\documentclass[12pt]{article}
\usepackage{amssymb,epsf}
\makeatletter
\if@titlepage
  \renewcommand\maketitle{\begin{titlepage}%
  \let\footnotesize\small
  \let\footnoterule\relax
  \null\vfil
  \vskip 60\p@
  \begin{center}%
    {\Large \bfseries\@title \par}%
    \vskip 3em%
    {\normalsize
     \lineskip .75em%
      \begin{tabular}[t]{c}%
        \@author
      \end{tabular}\par}%
      \vskip 1.5em%
    {\normalsize \@date \par}
  \end{center}\par
  \@thanks
  \vfil\null
  \end{titlepage}%
  \setcounter{footnote}{0}%
  \let\thanks\relax\let\maketitle\relax
  \gdef\@thanks{}\gdef\@author{}\gdef\@title{}}
\else
\renewcommand\maketitle{\par
  \begingroup
    \renewcommand\thefootnote{\fnsymbol{footnote}}%
    \def\@makefnmark{\hbox to\z@{$\m@th^{\@thefnmark}$\hss}}%
    \long\def\@makefntext##1{\parindent 1em\noindent
      \hbox to1.8em{\hss$\m@th^{\@thefnmark}$}##1}%
    \if@twocolumn
      \ifnum \col@number=\@ne
        \@maketitle
      \else
        \twocolumn[\@maketitle]%
      \fi
    \else
      \newpage
      \global\@topnum\z@   
      \@maketitle
    \fi
    \thispagestyle{plain}\@thanks
  \endgroup
  \setcounter{footnote}{0}%
  \let\thanks\relax
 \let\maketitle\relax\let\@maketitle\relax
  \gdef\@thanks{}\gdef\@author{}\gdef\@title{}}
\def\@maketitle{%
  \newpage
  \null
  \vskip 1em%
  \begin{center}%
    {\Large \bfseries\@title \par}%
    \vskip 1.5em%
    {\normalsize
      \lineskip .5em%
      \begin{tabular}[t]{c}%
        \@author
      \end{tabular}\par}%
    \vskip 1em%
    {\normalsize \@date}%
  \end{center}%
  \par
  \vskip 1.5em}
\fi
\renewcommand\section{\@startsection{section}{1}{\z@}%
                                     {-3.25ex\@plus -1ex \@minus -.2ex}%
                                     {1.5ex \@plus .2ex}%
                                     {\reset@font\normalsize\bfseries}}
\renewcommand\subsection{\@startsection{subsection}{2}{\z@}%
                                    {3.25ex \@plus1ex \@minus.2ex}%
                                    {-1em}%
                                    {\reset@font\normalsize\bfseries}}
\renewcommand\subsubsection{\@startsection{subsubsection}{3}{\z@}%
                                    {3.25ex \@plus1ex \@minus.2ex}%
                                    {-1em}%
                                    {\reset@font\normalsize\it}}
\@addtoreset{equation}{section}
\newcommand{\note}[1]{\raisebox{1ex}{{\footnotesize \sf #1}}}
\newcommand{\rnote}[1]{\raisebox{1ex}{{\hspace*{-3mm} \scriptsize\sf#1}}
                       \hspace*{-4mm}}

\makeatother
\def\be{ \begin{equation}}          \def\ee{ \end{equation}}
\def\ba{ \begin{eqnarray}}          \def\ea{ \end{eqnarray}}
\def\nn{\nonumber}                  
\def\ve{\varepsilon} 

 \def\Z{\mathbb{Z}} \def\R{\mathbb{R}}
\def\mathN{\mathbb{N}}
\def\ra{\rightarrow} 
\def\o{\otimes}                     \def\bz{\bar z}

\def\bT{{\overline T}} \def\bW{{\overline W}}
\def\bcW{{\overline {\cal W}}} \def\bJ{{\overline J}}
\def\bJ{{\overline J}} \def\sW{{\sf W}} 
\def\sJ{{\sf J}}  \def\sT{{\sf T}} 
\def\baa{{\bar a}}
\def\tr{\mbox{\it tr\/}} 
\def\oh{\frac{1}{2}}

\def\cV{{\cal V}}   \def\cH{{\cal H}}   \def\cU{{\cal U}}
\def\cW{{\cal W}}   \def\cM{{\cal M}}    
\def\a{\alpha }          \def\b{\beta }          
\def\d{\delta }               \def\ve{\varepsilon}
           \def\s{\sigma}
\font\fatma=cmbxti12
\title{Boundary Deformation Theory \\[2mm] and \\[2mm] 
       Moduli Spaces of D-Branes\\[2mm] \phantom{b}}
\author{Andreas Recknagel \rnote{$\;$1$*$} \ \ \ and \ \ 
 Volker Schomerus \rnote{$\;$2} 
\\[7mm] 
\note{1} Max-Planck-Institut f\"ur Mathematik in den Naturwissenschaften
\\ Inselstra\ss e 22-26, D--04103 Leipzig, Germany
\\[3mm]
\note{2} II. Institut f\"ur Theoretische Physik, Universit\"at Hamburg,
\\ Luruper Chaussee 149, D--22761 Hamburg, Germany}

\date{}
\begin{document}
\begin{titlepage}      \maketitle       \thispagestyle{empty}

\begin{abstract}
\noindent 
Boundary conformal field theory  is the suitable framework  
for a microscopic treatment of D-branes in arbitrary CFT backgrounds. 
In this work, we develop boundary deformation theory in order to 
study the changes of boundary conditions generated by marginal 
boundary fields. The deformation
parameters may be regarded as continuous moduli of D-branes. 
We identify a large class of boundary fields which are shown 
to be truly marginal, and we derive closed formulas describing 
the associated deformations to all orders in perturbation theory.
This allows us to study the global topology properties of the 
moduli space rather than local aspects only.
As an example, we analyse in detail the moduli space of $c=1$ 
theories, which displays various stringy phenomena. 
\end{abstract}
\vspace*{-18cm}
{\tt {DESY 98-185 \hfill MIS-preprint\ No.\ 60}}\\
{\tt {hep-th/9811237 \hfill}}
\vfill
\noindent\phantom{wwwx}{\small e-mail addresses: }
{\small\tt anderl@mis.mpg.de, vschomer@x4u.desy.de} 

\smallskip
\noindent\phantom{wwwx}{\small ${}^*\;$Address after March 1, 1999: 
Max-Planck-Institut f\"ur Gravitationsphysik,}
\newline
\noindent\phantom{wwwx{\small ${}^*\;$}}
{\small Albert-Einstein-Institut, Schlaatzweg 1, D--14473 Potsdam, Germany}
\end{titlepage}

\thispagestyle{empty} 
\setcounter{page}{0} 
\phantom{b} 
\newpage
\section{Introduction} 

Since Polchinski's discovery that D-branes \cite{DLP} provide 
a string realization of supergravity solitonic $p$-branes 
in \cite{Pol1,Pol2,Pol3}, non-perturbative  
effects have become accessible within string theory. This has changed 
the perspective of both string theory and gauge theories drastically. 
In particular, a net of dualities has emerged relating different field or  
string theories in the unified picture of $M$-theory \cite{Wit1}; see e.g.\ 
\cite{Ler,GiKu,Bac1,Sen5} for reviews and further references. 
More recently, this has led to conjectures of rather direct equivalences 
between string and supergravity theories on one side and 
gauge theories on the other \cite{Mal2,GKP,Wit3}.

D-branes are the most important new objects in this development. They have 
mainly been investigated from a target geometry and classical field 
theory point of view, where they appear as ``defects'' of various dimensions
to which closed strings can couple and which support gauge theories. 
In the flat background case, there exists a well-known alternative  
world-sheet approach using the boundary state formalism \cite{CLNY,PoCa}; it 
provides an effective handle on explicit string calculations but also allows 
to reproduce the classical behaviour of D-branes in the low-energy limit, 
see e.g.\ \cite{Bac,Li,CaKl,GrGu2,BG1,BCV,VFPSLR}. This formulation was 
extended somewhat beyond the flat case e.g.\ in \cite{HINS2}, see also 
\cite{OOY,Stan}, but to give a fully general formulation of D-branes 
in arbitrary CFT backgrounds \cite{ReSc,FuSc2} with no a priori classical 
counterpart requires more refined techniques. Those are provided by 
conformal field theory on surfaces with boundaries as developed mainly 
by Cardy \cite{Car1,Car2,Car3,CaLe} and first introduced into string 
theory by Sagnotti \cite{Sag1,BiSa1,Sag2}.

Techniques from conformal field theory are particularly well 
developed for rational models in which the state space 
decomposes into a finite number of sectors of some chiral 
symmetry algebra. This general remark applies to boundary 
theories in particular and means that boundary conditions 
with a large symmetry are the easiest to construct. 
In fact, for a certain class of rational models, 
Cardy managed to write down universal solutions \cite{Car3}. 
A variant of Cardy's ideas was used in \cite{ReSc} to obtain 
boundary conditions that describe D-branes in Gepner models. 
The set of such rational boundary theories is typically 
discrete. 

Continuous moduli, therefore, are an important feature of strings 
and branes that is rather difficult to handle with the algebraic 
techniques of CFT. Here, geometry and gauge theory undoubtedly 
are more efficient in producing quick results. Still, there are 
reasons to try and investigate moduli spaces within the
CFT approach: First of all, it is one of the fundamental ideas 
of string theory to treat space-time as a derived concept, not as 
part of the input data. Moreover, when starting a discussion of 
string or brane moduli spaces from geometrical notions, one runs the 
risk of missing some of the non-classical features of the moduli space 
and of the dynamics of massless fields. Finally, the efficiency of 
geometric approaches to moduli very much depends on the background and on
space-time supersymmetry; CFT methods, on the other hand, not only are 
background independent but also more robust when  
the amount of supersymmetry is reduced.

Within the CFT setting, moduli are the parameters of deformations 
generated by marginal operators -- more specifically, of marginal 
{\em boundary} perturbations if one is interested in D-brane 
moduli. Up to now, there does not seem to exist 
a systematic treatment of marginal deformations of boundary CFTs 
in the literature. There are, however, interesting case studies partly 
motivated by open string theory \cite{BiSa2,BPS,GrGu1}, partly by 
dissipative quantum mechanics \cite{CFF,CaKle,CKLM,CKMY,PoTh0,PoTh}.

The present paper aims at closing this gap and at presenting a 
general treatment of marginal perturbations of conformal boundary 
conditions.  
A careful analysis of the properties of marginal operators 
reveals that there is a large class of deformations which 
can be treated to all orders in perturbation theory. 
For deformations of CFTs on the plane, this is possible 
only for very few cases so that, usually, only local properties 
of the closed string moduli space are accessible from CFT. In contrast, 
the closed formulas we obtain for marginal boundary deformations 
allow us to recover global topological aspects of the D-brane moduli 
space from CFT.

{}From the $\s$-model interpretation one expects that continuous 
brane moduli should reveal some information about the underlying 
{\em target} space itself, the simplest geometric moduli being the position 
coordinates of D-branes in the target.  And indeed, we shall see 
target geometry  -- ``blurred'' and enriched by stringy effects --
emerging from our CFT analysis even though our 
starting point is purely algebraic with no initial reference to a 
classical $\sigma$-model description.

The simplest class of deformations we consider are the so-called 
{\em chiral deformations}. Roughly speaking, branes obtained from 
each other by chiral deformations are related through continuous
symmetries of the target space. {\em Non-chiral} deformations, 
however, are capable of moving branes between inequivalent positions 
not related by any continuous symmetry. In particular, 
they can push the brane into some singularity of the underlying 
target space (e.g.\ a fixed point of some orbifold group). The 
geometric singularity becomes manifest within the CFT description 
through a breakdown of certain sewing relations 
resp.\ the {\em cluster property} to be discussed below. 
In addition, we shall encounter some non-chiral deformations 
without an immediate target interpretation. 

\bigskip

The paper is organized as follows:
Section 2 introduces some tools from boundary conformal field 
theory needed throughout the text. It is also designed so as 
to make the presentation self-contained. 
In the end, we will explain the cluster property mentioned above 
and introduce the 
notion of a ``self-local''  boundary field that will become 
a crucial ingredient in our discussion of D-brane moduli spaces. 

In Section 3, we will give a detailed general discussion 
of marginal boundary deformations. We will show that whenever 
a marginal boundary operator is self-local it is truly 
marginal to all orders in the perturbation parameter (Subsection 3.2).  
Moreover, we present formulas which allow to compute structure constants 
of the deformed theory to all orders in perturbation theory. 
For reasons to become clear later, deformations generated by 
self-local boundary fields will also be called ``analytic''.  

Currents from the chiral symmetry algebra are special cases 
of self-local marginal fields; they generate group manifold pieces within 
the moduli space, and the corresponding deformed models can be described 
through simple closed formulas (Subsections 3.3-4). 
Subsection 3.5 contains further observations on the effect of (non-chiral) 
analytic deformations on Ward identities and spectrum of boundary 
excitations. In particular, we shall see which symmetries remain 
unbroken and which part of the brane partition function 
is independent of the strength of the perturbation. This 
explains and generalizes observations made for deformations 
of free bosonic boundary theories in \cite{CKLM}. 

Section 4 contains a more or less complete analysis of truly marginal 
boundary deformations of $c=1$ theories, which provide explicit examples
for all elements of our general construction. During the discussion, 
which subsumes the material of \cite{CKLM,PoTh} and leads to new 
results on orbifold models, we shall see that quantum field theoretical 
``subtleties'' like the cluster property are crucial in determining 
the topology of the moduli space of boundary conditions. 

A summarizing description of this $c=1$ brane moduli space is given in 
Section 5, with emphasis put on its non-classical features. Some of 
these are familiar effects from stringy geometry, while the interpretation 
of others remains to be found. We
conclude the paper with a brief outlook on possible extensions and on 
applications of our framework to the investigation of D-brane moduli 
spaces in arbitrary backgrounds. 
\medskip

We hope that our methods will also be useful for condensed 
matter problems, which represent the second important field of 
application of boundary CFT. We have already mentioned investigations  
of boundary perturbations in connection with dissipative quantum 
mechanics. The influence of dissipation on a particle in an infinite
periodic potential is described by the boundary sine-Gordon model 
which at the same time appears to be closely related to the Kondo problem 
\cite{FLS}. The latter deals with the marginally relevant perturbation 
induced by an impurity spin in a magnetic alloy, see e.g.
\cite{AfLu1,Aff,Lud} and references therein.

\section{Boundary conditions in conformal field theory} 

In this section we present a brief survey of boundary conformal 
field theory and fix the notations
used throughout the paper. It is explained in some detail
how boundary theories are parameterized by the choice of {\em 
gluing maps} $\Omega$ and the {\em structure constants} $A^\a_\varphi$
appearing in the 1-point functions for bulk fields of the theory. 
The last subsection is devoted to boundary fields. In particular, 
we introduce a notion of {\em locality} that will become crucial for 
the deformation theory to be developed below.  

\subsection{The bulk conformal field theory.} 

All constructions of boundary conformal field theories start 
from a usual conformal field theory on the complex plane, 
which we shall refer to as {\em bulk theory}. It consists 
of a space $\cH^{(P)}$ of states equipped with the action of a 
Hamiltonian $H^{(P)}$ and of field operators $\varphi(z,\bar z)$, 
which can be assigned uniquely to elements in the state space $\cH^{(P)}$
via the state-field correspondence, i.e.\ 
\be \varphi(z,\bar z) \ = \ \Phi^{(P)}(|\varphi\rangle ;z,\bar z) \ \ \ 
   \mbox{ for all } \ \ \ |\varphi\rangle \in \cH^{(P)}\ \ . 
\label{stfldcor}
\ee
The reverse relation is given by $\varphi(0,0) | 0\rangle 
= |\varphi \rangle$ where $|0\rangle$ denotes the vacuum 
state in $\cH^{(P)}$. 
\newline
The CFT is completely determined once we know all possible 3-point 
functions, or, equivalently, the coefficients of the operator 
product expansions (OPEs) for all fields in the theory. 
This task is often tractable since fields and states 
can be organized into irreducible representations of the 
observable algebra generated by the energy-momentum tensor and other 
chiral fields \cite{BPZ}. 
\medskip 

Chiral fields depend on only one of the coordinates $z$ or $\bar z$ 
so that they are either holomorphic, $W = W(z)$, or anti-holomorphic, 
$\bW = \bW(\bar z)$. The (anti-)\-holomorphic fields of a given 
bulk theory, or their Laurent modes $W_n$ and $\bW_n$ defined through 
\be    W(z) \ = \ \sum \, W_n \ z^{-n-h} \ \ \ , \ \ \ 
  \bW(\bar z) \ = \ \sum \, \bW_n \ \bar z^{-n-\bar h} \ \ , 
\label{modexp}
\ee  
generate two commuting {\em chiral algebras}, $\cW$ and $\bcW$. 
The Virasoro fields $T$ and $\overline T$ with modes $L_n$ and 
$\overline{L}_n$ are among the chiral fields of a CFT and, above,  
$h$ and $\bar h$ are 
the (half-) integer conformal weights of $W$ and $\bW$ wrt.\ $L_0$ and 
$\overline{L}_0$.  {}From now on we shall assume the two chiral 
algebras $\cW$ and $\bcW$ to be isomorphic. 

The state space of a CFT on the plane admits a 
decomposition $\cH^{(P)} = \bigoplus_{i,j}  \cV^i \o \cV^j$ 
into irreducible representations of the two commuting chiral algebras.
$\cV^0$ refers to the vacuum representation  -- 
which is mapped to $\cW$ via the state-field correspondence $\Phi^{(P)}$. 
\newline
The irreducible representations $\cV^i$ of $\cW$ 
acquire a (half-)integer grading under the action of $L_0$  
so that they may be decomposed as $\cV^i = \bigoplus_{n\geq 0} V^i_n$. 
We assume that the $V^i_n$ are finite-diemensional. 
Let $V^i_0 \subset \cV^i$ 
be the eigenspace of $L_0$ with lowest eigenvalue. It carries an 
irreducible action of all the zero modes $W_0$. We will denote 
the corresponding linear maps by $X^i_W$,
\be    X^i_W \ := \ W_0\ |_{V^i_0} : V^i_0 \ \longrightarrow \ 
                 V^i_0 \ \ \ \mbox{ for all chiral fields $W$} 
\ \ . 
\label{Xdef}
\ee       
The whole irreducible representation $\cV^i$ may be recovered 
from the elements of the finite-dimen\-sional subspace $V^i_0$ 
by acting with $W_n,\ n < 0$.            

Using the state-field correspondence $\Phi^{(P)}$, we can assign 
fields to all states in $V^i_0 \o V^j_0$. We shall assemble them 
into a single object which one can regard as a matrix of fields 
after choosing some basis in the subspaces $V^i_0$ and $V^j_0$, 
\be \varphi_{ij}(z,\bar z) \ : = \ \Phi^{(P)}(V^i_0 \o V^j_0; z,
   \bar z)\,:\  V^j_0 \o \cH^{(P)} \ \longrightarrow \ V^i_0 \o 
   \cH^{(P)} \ \ . 
\label{phiij}
\ee
In case the $\cV^i$ are $W$-algebra highest weight representations, 
$\varphi_{ij}(z,\bar z)$ are simply all the (Virasoro primary) fields 
which arise from a $\cW$-primary through the action of $W$-algebra  
zero modes.

\subsection{Boundary theories and the gluing map.} 

With some basic notations for the (``parent'') bulk theory set up, we can 
begin our analysis of {\em associated} boundary theories (``descendants''). 
These are conformal field theories on the upper half-plane $\Im z \geq 0$ 
which, in the interior $\Im z>0$, are locally equivalent to the 
given bulk theory: The state space $\cH^{(H)}$ of 
the boundary CFT is equipped with the action of a Hamiltonian 
$H^{(H)}$ and of bulk fields $\varphi(z,\bar z)$
-- still well-defined for $\Im z >0$ --  assigned to the 
{\em same} elements $\varphi$ that were used to label fields in the 
bulk theory. Accordingly, we demand that all the OPEs of bulk fields 
coincide with the OPEs of the bulk theory.
\newpage
Note that, in general, the boundary theory contains a lot more bulk 
fields than it has states. We will see shortly which fields are 
in one-to-one correspondence to the states in $\cH^{(H)}$. 

\medskip

Considering all possible conformal boundary theories associated to a bulk 
theory whose chiral algebra is a true extension of the Virasoro 
algebra is, at present, too difficult a problem to be addressed 
seriously. For the moment, we restrict our considerations 
to that class of boundary conditions which leave the whole symmetry 
algebra $\cW$ unbroken. More precisely, we assume that all chiral 
fields $W(z), \bW(\bz)$ can be extended analytically to the 
real line and that there exists a local automorphism $\Omega$ 
-- called the {\em gluing map} -- of the chiral algebra $\cW$ 
such that \cite{ReSc} 
\be T(z) \ = \ \bT (\bar z)   \ \ \ \mbox{ and }\ \ \  
   W(z) \ = \ \Omega(\bW) (\bar z) \ \ \mbox{ for } \ \ z = \bar z\ \ . 
\label{gluecond}
\ee
The first condition simply forbids an energy flow across the 
boundary; it is included in the second equation if we require 
$\Omega$ to act trivially on the Virasoro field. 
Note also that $\Omega$ induces an automorphism $\omega$ 
of the fusion rule algebra.  
\medskip

Our assumption on the existence of the gluing map $\Omega$ has 
the powerful consequence that it gives rise to an action of 
one chiral algebra $\cW$ on the state space $\cH\equiv \cH^{(H)}$
of the boundary theory. To see this, we combine the chiral 
fields $W(z)$ and $\Omega \bW(\bar z)$ into a single object 
$\sW (z)$ defined on the whole complex plane such that    
$$   \sW(z) \ :=\ \left\{ \begin{array}{ll} 
     W(z) \ \ &\mbox{ for } \ \ \Im z \geq 0 \\[2mm]
     \Omega \bW(\bar z) \ \ &\mbox{ for } \ \ \Im z < 0  
     \end{array} \right.  \ \ . 
$$ 
Because of the gluing condition along the boundary, this field
is analytic and we can expand it in a Laurent series $\sW(z)  = 
\sum_n \, W^{(H)}_n z^{-n-h}$, thereby introducing the modes 
$W_n\equiv W_n^{(H)}$. 
These operators on the state space $\cH$ are easily seen to 
obey the defining relations of the chiral algebra $\cW$.
Note that there is just one such action of $\cW$ constructed
out of the two chiral fields $W(z)$ and $\Omega \bW(\bar z)$. 
\medskip

In the usual way, the representation of $\cW$ on $\cH$ leads 
to Ward identities for correlation functions of the boundary 
theory. They follow directly from the singular parts of the 
operator product expansions of the field $\sW$ with the bulk 
fields $\varphi(z,\bar z)$ which are fixed by our 
requirement of local equivalence between the bulk theory and 
the bulk of the boundary theory. To make this more precise,  
we introduce the notation $\sW_>(z) = \sum_{n > -h} W_n 
z^{-n-h}$ for the singular part of the field $\sW$. The 
singular part of the OPE is then given by  
\newpage
\ba 
{\lefteqn{\sW(w) \, \varphi(z,\bar z)  \sim   
[\, \sW_>(w)\, ,\, \Phi(\varphi;z,\bar z)\, ]}} \label{WOPE}\\[2mm] 
& = & \sum_{n > -h}  \left( \frac{1}{(w-z)^{n+h}} \, \Phi\bigl(W^{(P)}_n 
 \varphi; z, \bar z\bigr) + \frac{1}{(w-\bar z)^{n+h}} \, 
  \Phi\bigl(\Omega\bW{}^{(P)}_n \varphi; z,\bar z\bigr)\right) \ \ . \nn 
\ea   
Here, the symbol $\sim$ means that the right hand side gives only 
the singular part of the operator product expansion, and we have 
placed a superscript $(P)$ on the modes $W_n, \bW_n$ to 
display clearly that they act on the elements $\varphi \in 
\cH^{(P)}$ labeling the bulk fields in the theory (superscripts 
 $(H)$, on the other hand, will be dropped).
The sum on the right hand side of eq.\ (\ref{WOPE}) is always 
finite because $\varphi$ is annihilated by all modes 
$W{}^{(P)}_m, \bW{}^{(P)}_m$ with sufficiently large $m$. For $\Im w >0$ 
only the first terms involving $W{}^{(P)}_n$ can become singular 
and the singularities agree with the singular part of the 
OPE between $W(w)$ and $\varphi(z,\bar z)$ in the bulk theory. 
Similarly, the singular part of the OPE between $\Omega \bW(w)$ and 
$\varphi(z,\bar z)$ in the bulk theory is reproduced by the 
terms which contain $\bW{}^{(P)}_n$, if $\Im w < 0$.    
\smallskip

As it stands, the previous formula is rather abstract. So, 
let us spell out at least one more concrete example in which 
the chiral field $\sW$ has dimension $h=1$  (we shall denote 
such chiral currents by the letter $J$ from now on) and in 
which  the field $\varphi$ is replaced by the matrix $\varphi_{ij}$  
of fields that were assigned to states $\varphi \in V^i_0 \o V^j_0$
through eq.\ (\ref{phiij}). Since the latter are annihilated by all 
the modes $J_n, \bJ_n$ with $n > 0$,  equation (\ref{WOPE}) 
reduces to      
\ba 
\sJ(w) \, \varphi_{ij}(z,\bar z) \ \sim \  
 \frac{ X_J^i }{w-z} \, \varphi_{ij}(z,\bar z)  \ - \ 
 \varphi_{ij}(z,\bar z) \, \frac{X_{\Omega \bJ}^j}{w-\bar z}  \ \ .  
\label{JOPE} 
\ea
The linear maps $X_W^i$ and $X_{\Omega \bJ}^j$ were introduced 
in eq.\ (\ref{Xdef}) above; they act on $\varphi_{ij}: V^j_0
\o \cH \rightarrow V^i_0 \o \cH$ by contraction in the first
tensor component $V^i_0$ resp.\ $V^j_0$.         
\smallskip

Ward identities for arbitrary $n$-point functions of fields 
$\varphi_{ij}$ follow directly from eq.\ (\ref{WOPE}). They have 
the same form as those for chiral conformal blocks in a bulk CFT 
with $2n$ insertions of chiral vertex operators with charges 
$i_1,\dots,i_n,\omega(j_1),..,\omega(j_n)$, see e.g.\ 
\cite{Car1,Car2,ReSc,FuSc2}. In many concrete examples, 
one has rather explicit expressions for such chiral blocks. 
So we see that objects familiar from the construction of 
bulk CFT can be used as building blocks of correlators in 
the boundary theory (``doubling trick''). Note, however, that 
the Ward identities depend on the gluing map $\Omega$.

\subsection{One-point functions.} 
Using the Ward identities described in the previous subsection 
together with  the OPE in the bulk, we can 
reduce the computation of correlators involving $n$ bulk fields 
to the evaluation of $1$-point functions $\langle \varphi_{ij}
\rangle_\a$. They need not vanish in a boundary CFT because translation 
invariance along the imaginary axis is broken, and they may depend on the
possible boundary conditions $\a$ along the real line.

To control the remaining freedom, we notice that the transformation 
properties of $\varphi_{ij}$ with respect to $L_n,\ n= 0, \pm 1,$ 
and the zero modes $W_0$,  
\ba [\, W_0\, , \, \varphi_{ij}(z,\bar z)\, ] & = & 
    X_W^i\ \varphi_{ij}(z,\bar z) \ - \ \varphi_{ij}(z,\bar z) 
  \ X_{\Omega \bW}^j \ \ , \ \ \nn \\[3mm]   
[\, L_n\, , \, \varphi_{ij}(z,\bar z)\, ] & = & 
   z^n\, ( \, z \partial + h_i(n+1)\, ) \varphi_{ij}(z,\bar z)   
  \nn \\[2mm] & & \hspace*{5mm} +  
   \bar z^n\, ( \, \bar z {\overline \partial} + \bar h_i(n+1)\, ) 
    \varphi_{ij}  (z,\bar z) \ \  \nn 
\ea 
determine the 1-point functions up to scalar factors. Indeed, 
an elementary computation reveals that $\langle \varphi_{ij}
\rangle_\a$ must be of the form 
\be
 \langle \varphi_{ij} (z,\bar z) \rangle_\alpha  =  
 \frac{ A^\alpha_{ij}}{(z - \bar z)^{h_i + h_j}}\ \ 
 \label{1ptfct}  
\ee
where 
$$
 A^\a_{ij}: V^j_0 \rightarrow V^i_0  \quad\quad \mbox{ obeys } \quad
 \ \ \ X_W^i \ A^\a_{ij} \ = \ A^\a_{ij} \ X_{\Omega \bW}^j \ \ . 
$$
The intertwining relation in the second line implies $j = \omega^{-1}(i^+)$ 
as a necessary condition for a non-vanishing 1-point function ($i^+$ denotes
the representation conjugate to $i$), and 
thus we can put  $h_i + h_j= 2 h_i $ in the exponent in eq.\ (\ref{1ptfct})
because the gluing map acts trivially on the Virasoro field. 
Irreducibility of the zero mode representations on the subspaces $V^i_0$ 
and Schur's lemma imply that each matrix 
$A^\a_{ij}$ is determined up to one 
scalar factor. Hence, if there exist several boundary conditions
associated with the same bulk theory and the same gluing map 
$\Omega$, they can differ only by these scalar parameters in 
the 1-point functions. Once we know their values, we have 
specified the boundary theory. In particular, one can express 
the partition function $Z_{(\Omega, \a)}(q)$ of the 
theory in terms of the coefficients $A^\a_{ij}$ (see eqs.\ 
(\ref{partfct},\ref{AeqBB}) below for precise formulas). 
\begin{figure} 
{\epsfbox{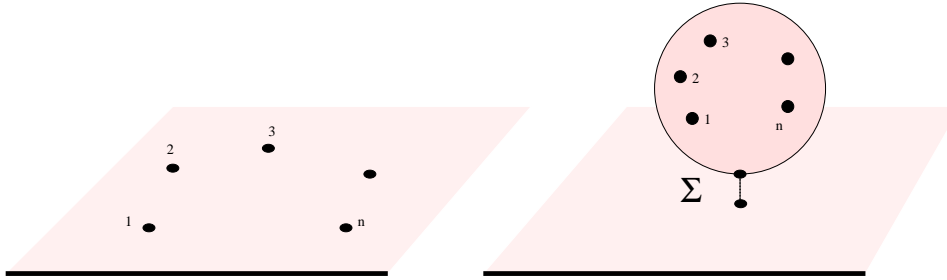}} \vspace*{4mm}
\caption{ \small With the help of operator product expansions in 
the bulk, the computation of $n$-point functions in a 
boundary theory can be reduced to computing 1-point 
functions on the half-plane. Consequently, the latter 
must contain all information about the boundary 
condition.} 
\end{figure}  
\medskip

The parameters in the 1-point functions are not completely 
free. In fact, there exist strong {\em sewing constraints} 
on them that have been worked out by several authors 
\cite{CaLe,Lew,PSS1,PSS,BPPZ}. The basic relation can be derived from the 
following {\em cluster property} of correlation functions: 
\ba \lefteqn{\hspace*{-10mm} \lim_{a \ra \infty} \ \langle 
    \varphi_1(z_1,\bz_1) \dots 
   \varphi_{P-1}(z_{P-1},\bz_{P-1}) \varphi_{P} (z_{P}+a, 
   \bz_P+a) \dots \varphi_N(z_N+a,\bz_N+a)\rangle} \nn \\[2mm] 
    & = &   \langle \varphi_1(z_1,\bz_1) \dots 
   \varphi_{P-1}(z_{P-1},\bz_{P-1})\rangle \ \langle 
   \varphi_{P} (z_{P},\bz_P) \dots \varphi_N(z_N,\bz_N)\rangle 
    \ \ . \label{cluster}\ea 
Here, $a$ is a real parameter, and the fields 
$\varphi_\nu =\varphi_{i_\nu,\bar\imath_\nu}$ on the 
right hand side can be placed at $(z_\nu,\bz_\nu)$ since the 
whole theory is invariant under translations parallel to the 
boundary. If the cluster property is combined with the Ward 
identities to evaluate 2-point functions of bulk fields, 
one obtains a constraint of the form
\be A^\a_i \, A^\a_j \ = \ \sum_k \ \Xi^{k}_{ij} \ 
   A^\alpha_0 \ A^\a_k 
   \quad \ \ \mbox{ with } \ \ \ A^\a_l \ = 
   \ A^\a_{l \omega(\bar l)} \ \ .  \label{class} \ee
It holds whenever the vacuum representation ``0'' occurs in the 
fusion product of $i$ with $\omega(\bar\imath)$ and of $j$ with 
$\omega(\bar\jmath)$. 
The coefficient $\Xi^k_{ij}$ can be expressed as  a combination 
of the coefficients in the bulk OPE and of the fusing matrix. 
In some cases, this combination has been shown to agree 
with the fusion multiplicities or some generalizations thereof 
(see e.g.\ \cite{PSS,FuSc1,FuSc3,BPPZ}). The importance of eq.\ 
(\ref{class}) for a classification of boundary conformal field 
theories has been stressed in a number of publications recently 
\cite{FuSc1,BPPZ,FuSh4} and is further supported by their 
close relationship with algebraic structures that entered 
the classification of bulk conformal field theories already 
some time ago (see e.g.\ \cite{Pas,ZuPe1,ZuPe2}). 

Let us remark that, from the string theory point of view, 
the 1-point functions give the couplings of closed string modes 
to a D-brane, i.e.\ generalized tensions and RR-charges. 
Eq.\ (\ref{class}) provides an example of non-linear constraints 
imposed on these couplings. 
\smallskip

In our discussion of boundary perturbations, we shall always 
depart from a set of correlation functions satisfying relation  
(\ref{cluster}). Anticipating a more detailed discussion 
below, we stress that boundary perturbations do not preserve
this property in general. One can often interpret the breakdown of 
rel.\ (\ref{cluster}) as a signal for the theory to develop a 
mixture of different ``pure'' (i.e.\ clustering) boundary 
conditions. Such phenomena are certainly expected to occur 
upon boundary perturbation and we will present some concrete 
examples later on.

\subsection{Boundary fields.} 
The action of $\cW$ on the state space of the boundary theory 
induces a decomposition $\cH= \bigoplus_i \cV^i$ (possibly 
with multiplicities) into irreducibles of $\cW$. It also implies
that the partition function may be expressed as a sum of 
characters $\chi_i(q)$ of the chiral algebra, 
$$     Z_{(\Omega,\a)}(q) \ :=  \ \tr_\cH (q^{L_0-c/24}) \ = \ 
      \sum_i n^{\Omega\a}_i \,
      \chi_i(q) \ \ \mbox{ where } \ \ n^{\Omega\a}_i \ \in \ 
       \mathN\ \ . 
$$  
There exists a one-to-one state-field correspondence 
$\Phi\equiv\Phi^{(H)}$ between states $\psi \in \cH$  
and so-called boundary fields $\psi(x)$ which are 
defined (at least) for $x$ on the real line \cite{ReSc}. The conformal 
dimension of a boundary field $\psi(x)$ can be read off from 
the $L_0$-eigenvalue of the corresponding state  $\psi \in \cH$. 
The boundary fields assigned to elements in the vacuum sector $\cV^0$
coincide with the chiral fields in the theory, i.e.\ 
$\sW(x) = \Phi(w;x)$ for some $w \in \cV^{0}$ and $\Im x = 0$. 
These fields can always be extended beyond the real line
and coincide with either $W$ or $\Omega \bW$ in the bulk. 
If other boundary fields admit such an extension, this
suggests an enlargement of the chiral algebra in the bulk 
theory.
\medskip

Following the standard reasoning in CFT, it is easy to conclude that 
the bulk fields $\varphi_{ij}(z,\bar z)$ give singular contributions 
to the correlation functions whenever $z$ approaches the real line. 
This can be seen from the fact that the Ward identities  describe a 
mirror pair of chiral charges $i$ and $\omega(j)$ placed on both sides 
of the boundary. 
Therefore, the singularities in an expansion of $\varphi(z,\bar z)
\equiv \varphi_{ij}(z,\bar z)$ around $x = {\Re} z$ are given
by primary fields which are localized at t $x$ on the 
the real line, i.e.\ the boundary fields $\psi(x)$. In other words,
the observed singular behaviour of bulk fields $\varphi(z,\bar z)$ near
the boundary may be expressed in terms of a bulk-boundary OPE 
\cite{CaLe} 
\be     \varphi(z,\bar z) \ = \ \sum_k \ (2 y)^{h_k - h - \bar h} 
        \ C^{\,\alpha}_{\varphi\; k}\  \psi_k(x)    \ \ . 
\label{bbOPE}
\ee
Here, $\psi_k(x)$ are primary fields of conformal weight $h_k$,
and $z= x+iy$. Which $\psi_k$ can possibly appear on the rhs.\ 
of (\ref{bbOPE}) is determined by the chiral fusion of $i$ and $\omega(j)$, 
but some of the coefficients $C^{\,\alpha}_{\varphi\, k}$ may vanish for 
some $\alpha$. One can show 
that $ C^{\,\alpha}_{\varphi\; {0}} = A^\alpha_\varphi /A^\alpha_0$; 
moreover, the $C^{\,\alpha}_{\varphi\; k}$ are related to the 1-point 
functions by generalizations of the constraints (\ref{class}), see 
e.g.\ \cite{Lew,PSS}.

In boundary conformal field theory one also considers boundary 
fields which induce transitions (``jumps'') between different boundary 
conditions $\a,\b$, see e.g.\ \cite{Car3}. 
These ``boundary condition changing operators'' are 
associated with vectors in a state space $\cH_{\a\b}$ depending 
on both boundary conditions, and they cannot be obtained 
from bulk fields through a bulk-boundary OPE. Even though 
we shall only consider homogeneous perturbations of boundary 
conditions which are constant all along the boundary, we  
will meet some boundary condition changing operators eventually: 
At certain values of the deformation parameter $\lambda$, it 
may happen that a perturbed theory describes a mixture (or 
superposition) of different clustering boundary conditions. 
In such cases, no jump is visible along the real axis, 
but there exist boundary fields which induce transitions 
between the various ``pure'' boundary conditions. 

\medskip

Having introduced the boundary fields $\psi(x)$, it is natural 
to extend the set of correlation functions and to consider 
correlators in which a number of boundary fields
$\psi_\nu(x_\nu) = \Phi(\psi_\nu,x_\nu)$ are inserted along with  
bulk fields: 
\be
 \langle \psi_1(x_1) \dots \psi_M(x_M) \, \varphi_1(z_1,\bz_1) \dots 
  \varphi_N(z_N,\bz_N) \rangle_\a \ \ \ \mbox{ for } \ \ \ 
  x_\nu < x_{\nu+1}\ \ \ . 
\label{bcorr} \ee
These functions are analytic in the variables $z_\nu$ throughout the
whole upper half-plane $\Im z_\nu > 0$. For the variables  $x_\nu$, 
the domain of analyticity is restricted to the interval $x_\nu \in \;
]x_{\nu-1}, x_{\nu+1}[$ on the boundary. In most cases, there exists 
no {\em unique} analytic continuation of $\psi_\nu(x_\nu)$ to other 
points on the real axis which lie beyond the insertion points 
of the neighbouring boundary fields. In fact, if we continue 
analytically along curves like the one shown in Figure 2, the result 
will typically depend on whether we move the field $\psi_\nu$ 
around $\psi_{\nu+1}$ in clockwise or anti-clockwise direction. 
There are certainly exceptions: Chiral fields $\sW(x)$, for 
example, do possess a unique analytic continuation to 
$x \in \R \setminus\{ x_\nu\}$.       

\begin{figure} 
{\epsfbox{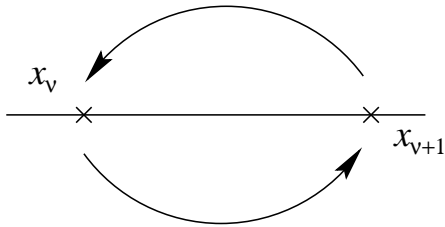}} 
\caption{ \small The curve $\gamma^\nu_{\nu+1}$ along which 
correlation functions are analytically continued to exchange 
the position of two neighboring boundary fields. In most cases
the result depends on the orientation of the curve.} 
\end{figure}  

Based on this discussion on analyticity, we want to introduce 
a notion of locality that will become important later on: Two 
boundary fields $\psi_1 (x_1)= \Phi(\psi_1;x_1)$ and $ \psi_2(x_1) 
= \Phi(\psi_2;x_2)$ are said to be {\em mutually local} if   
\be 
\Phi(\psi_1;x_1)\  \Phi(\psi_2;x_2) \ = \ \Phi(\psi_2;x_2)\  
  \Phi(\psi_1;x_1) \ \ \ \mbox{ with } \ \ x_1 < x_2\ \ . 
\label{local}
\ee
The equation is supposed to hold after insertion into arbitrary 
correlation functions, and for the right  hand side to make sense 
it is required that there exists a unique analytic continuation 
from $x_1< x_2 $ to $x_1 > x_2$.  
\newline
A boundary field $\psi(x) 
= \Phi(\psi;x)$ will be called {\em self-local} or {\em analytic} in 
the following if it is mutually {\em local with respect to itself}. 
(The second expression is chosen in view of the properties 
its correlation functions and of perturbations with self-local 
marginal operators.) 
\newline
Let us note that the OPE of two mutually 
local fields contains only pole singularities. In particular, 
the OPE of a self-local boundary field $\psi$ with conformal 
dimension $h_\psi = 1$ is determined up to a constant $K$ 
to be 
\be \label{locmargOPE}
  \psi(x_1) \, \psi(x_2) \ = \ \frac{K}{(x_1 - x_2)^2}\ +\  {\rm reg} \ \ 
 \ \ \mbox{ if } \ \ h_\Psi \, = \, 1 \ \ .  
\ee  
Boundary fields $\sW(x)$ from the chiral algebra are the simplest 
examples of analytic fields. They are not only local with respect to 
themselves but to all other boundary fields in the theory. 

It is crucial for our analysis of D-brane moduli to observe 
that further (non-chiral) analytic boundary fields $\psi$ can 
exist depending on the boundary condition under consideration. 
Unless they belong to some extended chiral symmetry (which 
means that the original chiral algebra $\cW$ was not chosen 
to be the maximal chiral symmetry), these self-local boundary 
fields $\psi$ will not possess a unique analytic continuation 
into the full upper half-plane. In fact, ``moving'' the boundary 
field $\psi$ around the insertion point of a bulk field $\varphi$
(by analytic continuation) 
can lead to a non-trivial monodromy in general. Whenever this 
happens, $\psi$ has no chance to be local wrt.\ all the 
boundary fields that appear in the bulk-boundary OPE of the 
bulk field $\varphi$. Consequently, a non-chiral analytic boundary 
field $\psi$ is only expected to be local wrt.\ a subset of boundary 
fields. The latter includes at least the chiral boundary fields
$\sW$ in addition to the field $\psi$ itself. 

The existence of non-chiral self-local fields is signaled by a 
partition function $Z_{\Omega\a}(q)$ that contains the vacuum 
character of a $W$-algebra $\cW_{\Omega\a}$ extending the chiral 
algebra $\cW \subset \cW_{\Omega\a}$ of the model. We shall see 
several examples in Section 4.

\section{Marginal perturbations of boundary conditions} 
  
Our aim is to study {\em  perturbations} (or {\em 
deformations}) of a boundary condition which are generated by 
{\em marginal} boundary fields. After some general remarks, we 
describe a class of perturbations -- which we call {\em 
analytic deformations} -- that are truly marginal to all 
orders in the perturbation expansion. They are induced by 
self-local boundary fields of dimension one. Therefore, 
deformations generated by chiral currents are among them 
and may serve us to illustrate the more general construction 
we propose. In the last subsection 
we investigate arbitrary non-chiral analytic deformations and derive 
some of their properties which hold to all orders 
in perturbation theory.

\subsection{The general prescription.} 
\def\la{\lambda}
Let us start from some boundary conformal field theory 
with state space $\cH = \cH^{(H)}_{(\Omega,\alpha)}$, where 
$(\Omega,\alpha)$ denotes the boundary condition along the 
real line.  Boundary operators $\psi(x) \in 
\Phi(\cH)$ may be used to define a new perturbed theory whose 
correlation functions are constructed from the unperturbed 
ones by the formal prescription 
\ba
\lefteqn{\langle\, \varphi_1(z_1,\bar z_1) \cdots 
    \varphi_N(z_N, \bar z_N)\,\rangle_{\alpha;\ \lambda\psi} 
  =   Z^{-1} \cdot \langle\, I_{\lambda\psi} \, \varphi_1(z_1,\bar z_1) \cdots 
    \varphi_N(z_N, \bar z_N) \,\rangle_{\alpha} }
\label{defcorra} \nn \\[2mm] 
 & := & Z^{-1} \sum_n \la^n \int \cdots \int_{x_i < x_{i+1}} 
       \frac{dx_1}{2\pi} \cdots \frac{dx_n}{2\pi} 
       \ \langle\, \psi(x_1) \cdots \psi(x_n) \ \varphi_1 \cdots 
        \varphi_N  \,\rangle_{\alpha}  \nn\\[2mm]
& = & Z^{-1} \sum_n \frac{\la^n}{n!}  \sum_{\s \in S_n}
       \int \cdots \int_{\raisebox{-3mm}{$ \scriptstyle 
        x_{\s(i)}\, <\, x_{\s(i+1)}$}} 
       \hspace*{-15mm} \frac{dx_1}{2\pi} \cdots \frac{dx_n}{2\pi} 
       \langle\, \psi(x_{\s(1)}) \cdots \psi(x_{\s(n)} ) 
     \  \varphi_1 \cdots 
    \varphi_N \,\rangle_{\alpha}  
      \nn   \ea
where $\lambda$ is a real parameter. The second sum in the lowest line 
runs over all elements in the permutation group $S_n$. 
Since all the $n!$ summands are identical, the last equality is 
obvious. It shows, however, that the symbol $I_{\lambda \psi}$ 
in the first line should be understood as a path ordered exponential
of the perturbing operator,  
\be  I_{\lambda\psi} \ = \ P \exp \bigl\lbrace\, \lambda 
   S_\psi\,\bigr\rbrace \ :=\  P \exp \bigl\lbrace\, 
    \lambda \int_{- \infty}^\infty \psi(x) \, 
     \frac {dx} {2\pi}\,\bigr\rbrace\ \  .  
\label{defcorrb}\ee
The normalization $Z$ is defined as the expectation value 
$Z = (A^\a_0)^{-1}\,\langle I_{\la \psi}\rangle_\a$. These expressions 
deal with deformations of bulk correlators only. If there are extra 
boundary fields present in the correlation function, the formulas 
need to be modified in an obvious fashion so that these boundary 
fields are included in the ``path ordering''. A particularly simple 
example of this type will be discussed shortly, but we refrain 
from spelling out the general formula here. 

To make sense of the above expressions (beyond the formal level), it is 
certainly necessary to regularize the integrals (introducing UV- and 
IR-cutoffs) and to renormalize couplings and fields (see e.g.\ 
\cite{CarL} for a discussion of bulk perturbations in 2D conformal 
field theory). IR divergences are usually cured by putting the 
system into a ``finite box'', i.e., in our case, by studying perturbations 
of finite temperature correlators; but this will not play 
any role below. On the other hand, we have to deal with UV 
divergences. So, let us introduce a UV cutoff $\ve$ such that 
the integrations are restricted to the region $|x_i - x_j|>\ve$. 
Thereby, all integrals become UV-finite before we perform the 
limit $\ve \rightarrow 0$.      
\medskip

In the following, we consider marginal boundary deformations 
where the conformal dimension $h$ of the perturbing operator
$\psi(x)$ is $h=1$ so that there is a chance to stay at the 
conformal point for arbitrary values of the real coupling $\lambda$ 
(we choose $\psi(x)$ to be anti-selfadjoint). 
If $h \neq 1$, the perturbation will automatically introduce a 
length scale and one has to follow the renormalization group 
(RG)-flow to come back to a boundary conformal field theory. For 
$h >1$, the perturbations are irrelevant so that one ends up with 
the original boundary theory. For $h < 1$, the perturbation is 
relevant and it is usually quite difficult to say precisely 
which conformal fix-point one reaches with a given relevant 
perturbation. Nevertheless, several non-trivial examples have been 
studied in the literature, see \cite{FSW,War2,Chim,DoPTWa,LeSa1,LeSaSi} 
and references therein, partially with the help of the thermodynamic 
Bethe ansatz. 

All these cases, however, share the common feature that (at the 
RG fix-point) the new conformal boundary theory is associated to 
the same bulk CFT -- since the local properties in the bulk are 
not affected by the ``condensate'' along the boundary. Thus, boundary 
perturbations can only induce changes of the boundary conditions. 
\smallskip

To begin our discussion of marginal perturbations with a 
boundary field $\psi(x)$, let us 
investigate the change of the two-point function $\langle \psi(x_1) 
\psi(x_2) \rangle_\a $ of the perturbing field $\psi$ itself  
under the deformation. Obviously, the first order contribution involves 
the following sum of integrals: 
$$ \int_{-\infty}^{x_1-\ve} \hspace*{-5mm} dx \,\langle \psi(x) 
   \psi(x_1) \psi(x_2) \rangle_\a 
 + \int_{x_1+\ve}^{x_2-\ve} \hspace*{-5mm} dx \,\langle \psi(x_1) 
     \psi(x) \psi(x_2) \rangle_\a
 + \int_{x_2-\ve}^{\infty} \hspace*{-5mm} dx \,\langle \psi(x_1) 
      \psi(x_2) \psi(x) \rangle_\a
 . $$
{}From the general form of the three point function
(with $x_1 < x_2 < x_3$)  
$$ \langle \psi(x_1) \psi(x_2) \psi(x_3)\rangle_\a =  
    \frac{C^\alpha_{\psi \psi \psi}}
       {(x_1-x_2)(x_1-x_3)(x_2-x_3)} \ ,
$$
it is easy to see 
that the first order contribution to the perturbation expansion 
is logarithmically divergent unless the structure constant 
$C^\a_{\psi\psi\psi}$ from the OPE of boundary fields 
vanishes. The divergence would 
force the conformal weight of the field $\psi$ away from the
initial value $h_\psi=1$ as we turn on the perturbation, i.e.\  
the marginal field $\psi$ is not truly marginal unless 
$C^\a_{\psi\psi\psi} = 0$. If there are 
several marginal boundary fields in the theory, degenerate 
perturbation theory gives a somewhat stronger condition: A
marginal field $\psi$ is {\em truly marginal only if $C^\a_{\psi\psi 
\psi'} = 0 $} for all marginal boundary fields $\psi'$ in the 
theory -- cf.\ e.g.\ \cite{DVV} for the bulk case. 
Eq.\ (\ref{locmargOPE}) shows that self-local marginal boundary 
operators do satisfy this first order condition. 
One should stress, however, that this is merely a 
necessary condition. Since it was derived within first order 
perturbation theory it is by no means sufficient to guarantee 
true marginality in higher orders of the perturbation series. 

\subsection{Truly marginal operators.}
The first order condition is how far general investigations of marginal 
{\em bulk} perturbations go.  
Our main aim here is to prove that every {\em self-local marginal boundary 
operator}  is indeed {\em truly marginal} to all orders and therefore 
generates a deformation of a boundary CFT. 
To this end, let us assume that 
the perturbing marginal field $\psi(x)$ is self-local in the sense discussed 
at the end of the previous section. Then the above expression for the 
deformed bulk correlation functions can be rewritten as
\ba  \label{bcfirst} 
\lefteqn{\langle\, \varphi_1(z_1,\bar z_1) \cdots 
    \varphi_N(z_N, \bar z_N)\,\rangle^{\ve}_{\alpha;\ \lambda\psi}} \\[2mm]
& = & Z^{-1} \sum_n \frac{\la^n}{n!}  
       \int_{-\infty}^\infty \cdots \int_{-\infty}^\infty 
        \frac{dx_1}{2\pi} \cdots \frac{dx_n}{2\pi} 
       \langle\, \psi(x_1) \cdots \psi(x_n ) 
     \  \varphi_1 \cdots 
    \varphi_N \,\rangle_{\alpha}  
      \nn   \ea
where all integrals are taken over the real line with the 
regions $|x_i - x_j| < \ve$ removed as before. Based on the  
OPE (\ref{locmargOPE}) of $\psi$, it is not difficult to see that 
the divergences in $\ve$ from the numerator cancel those from the 
denominator so that the limit $\ve \to 0$ of the deformed bulk field 
correlator can be taken. Moreover, as we are dealing with a self-local 
marginal operator, this limit can be 
written as 
\ba
\lefteqn{
\langle\, \varphi_1(z_1,\bar z_1) \cdots 
    \varphi_N(z_N, \bar z_N)\,\rangle_{\alpha;\ \lambda\psi}
= \lim_{\ve\to 0} \; \langle\, \varphi_1(z_1,\bar z_1) \cdots 
    \varphi_N(z_N, \bar z_N)\,\rangle^{\ve}_{\alpha;\ \lambda\psi}}
\label{analdef} \\[2mm]
& \phantom{xxxxx} = & \sum_n \frac{\la^n}{n!}  
       \int_{\gamma_1} \cdots \int_{\gamma_n}
        \frac{dx_1}{2\pi} \cdots \frac{dx_n}{2\pi} 
       \langle\, \psi(x_1) \cdots \psi(x_n ) 
     \  \varphi_1 \cdots 
    \varphi_N \,\rangle_{\alpha}  
      \nn   \ea  
where $\gamma_p$ is the straight line parallel to the real 
axis with $\Im \gamma_p = i \ve/p $,  and it can be computed 
through contour integration. The expression on the 
right hand side is manifestly finite, and it is independent 
of $\ve$ as long as $\ve <$ {\rm min\/}$(\Im z_i)$ where $z_i$ 
denote the insertion points of bulk fields. Thus, the above 
formula allows us to construct the perturbed bulk correlators 
to all orders in perturbation theory. In particular, it 
determines the deformation of bulk 1-point functions and 
hence the deformation of the structure constants 
$A^\a_{\varphi}$ which parameterize the possible boundary 
theories along with the gluing map.  
\smallskip

The extension of these ideas to the deformation of 
boundary correlators meets some obstacles. In fact, 
formula (\ref{bcfirst}) admits for the obvious 
generalization
\ba
\lefteqn{\langle\, \psi_1(u_1) \cdots \psi_M(u_M)\  
    \varphi_1(z_1,\bar z_1) \cdots 
    \varphi_N(z_N, \bar z_N)\,\rangle_{\alpha;\ \lambda\psi}} 
\label{bbcfirstdef}\\[2mm]
& = &  Z^{-1} \sum_n \frac{\la^n}{n!}  
       \int_{-\infty}^\infty \cdots \int_{-\infty}^\infty
        \frac{dx_1}{2\pi} \cdots \frac{dx_n}{2\pi} \;
       \langle\, \psi(x_1) \cdots \psi(x_n ) \;\psi_1 
       \cdots \psi_M
     \;  \varphi_1 \cdots 
    \varphi_N \,\rangle_{\alpha}  
      \nn   
\ea  
{\em if and only if} the boundary fields $\psi_1,\ldots,\psi_M$ are 
{\em local} with respect to the perturbing field $\psi$. As we have 
argued in the previous section, this is usually a strong 
constraint on boundary fields. The integrals on the rhs.\ of 
eq.\ (\ref{bbcfirstdef}) diverge as $\ve \rightarrow 0$ whenever 
the iterated OPE of the perturbing field $\psi$ with one of 
the boundary fields $\psi_i$ contains poles of even order. The 
(renormalized) correlation functions are again obtained 
through contour integration,  
\ba
\lefteqn{\langle\, \psi_1(u_1) \cdots \psi_M(u_M)\  
    \varphi_1(z_1,\bar z_1) \cdots 
    \varphi_N(z_N, \bar z_N)\,\rangle_{\alpha;\ \lambda\psi}} 
\label{banaldef}\\[2mm]
& = &  \sum_n \frac{\la^n}{n!}  
       \int_{\gamma_1} \cdots \int_{\gamma_n}
        \frac{dx_1}{2\pi} \cdots \frac{dx_n}{2\pi} \;
       \langle\,  \psi(x_1) \cdots \psi(x_n ) \; \tilde \psi_1 
       \cdots \tilde \psi_M
     \;  \varphi_1 \cdots 
    \varphi_N \,\rangle_{\alpha}  
      \nn   
\ea
where the fields $\tilde \psi_i$ in the correlator on the 
right hand side are given by
$$ \tilde \psi_i\ = \ \Bigl[ e^{\frac12 \lambda \psi}\, \psi_i 
   \Bigr](u_i) := \sum_{n=0}^{\infty} {\lambda^n\over 2^n n!}\; 
   \oint_{C_1} \!{dx_1\over 2\pi} \cdots \oint_{C_n} 
    \!{dx_n\over 2\pi} \; \psi_i(u_i)\psi(x_n) \cdots \psi(x_1)
 \ \  , 
$$
and $C_\nu$ are small circles around the insertion point 
of $\psi_i$.  Since the contour integrals on the rhs.\ pick 
out simple poles, the fields  $\psi_i$ and $\tilde \psi_i$ 
have the same conformal dimension -- $\tilde \psi_i$ can be 
regarded as the image of $\psi_i$ under a ``rotation'' generated 
by the perturbing field $\psi$.   

With the help of eq.\ (\ref{banaldef}) we are able to study  
the deformation of $n$-point functions of the (self-local) perturbing 
field $\psi$ itself. Notice that the OPE (\ref{locmargOPE}) 
contains no first order poles so that the fields  $\tilde \psi$ 
and $\psi$ coincide; in fact, all the contour integrals in eq.\ 
(\ref{banaldef}) are zero if there is no bulk field inserted 
in the upper half-plane. Hence, any perturbative correction to the 
$n$-point function of $\psi$ vanishes -- which implies that self-local 
marginal field are truly marginal. 

\subsection{Chiral marginal boundary perturbations.} 
For the time being, let us restrict to perturbations with local 
boundary fields $\sJ$ taken from the chiral algebra, i.e.\ we 
shall analyze perturbations generated by fields assigned to 
elements in the subspace $\cV^0_1 \subset \cH$. Such fields 
are local wrt.\ all bulk and boundary fields, so that eq.\ 
(\ref{banaldef}) may be applied to correlators involving 
arbitrary bulk and boundary fields. Consequently, a complete 
non-perturbative picture of the deformation can be given, 
including a proof of the invariance of the partition 
function.   

\subsubsection{Deformation of the gluing map.} 
\def\ve{\varepsilon} \def\ra{\rightarrow} \def\vac{|0\rangle}
Our first goal is to describe the effect a marginal 
perturbation with the boundary current $\sJ$ has on 
the {\em gluing map} $\Omega$. To this end, we phrase the 
content of the gluing condition (\ref{gluecond}) as follows: Suppose we insert 
the field $W(z+2i\d) - \Omega\bW(\bar z - 2i\d)$ with $z = 
\bar z$ into an arbitrary correlation function of the 
unperturbed theory. Then, by taking the limit $\d \rightarrow 
0^+$, we move $W$ and $\Omega\bW$ to the boundary until the 
correlator vanishes at $\d = 0$. Now we want to understand 
how the presence of the perturbation $P \exp ( \lambda S_J)$ 
influences this situation. In more formal terms, we need to 
evaluate the expression  
\ba \lefteqn{ 0  =  \lim_{\d \ra 0} \ P e^{ \lambda S_J} \ \ \left( W(z_\d) - 
             \Omega\bW (\bar z_\d) \right)} \nn \\[2mm]
  & = &  \lim_{\d \ra 0}\
         \sum_{n=0}^{\infty} \frac{\lambda^n}{n!} \int_{\gamma_1}  
        \cdots \int_{\gamma_n} \!\frac{dx_1}{2\pi}\cdots\frac{dx_n}{2\pi} \;
         \sJ(x_1) \cdots \sJ(x_n) \, \left( W(z_\d) - 
         \Omega\bW (\bar z_\d) \right)\nn 
\ea
where we have used $z_\d = z+2 i\d$, $\bar z_\d =z-2 i \d$ and inserted the 
definition of the operator $ P\exp (\lambda S_J)$ underlying formula 
(\ref{analdef}). Our next step involves closing 
the integration contours $\gamma_i$ either in the upper or in the 
lower half-plane. Let us  choose the upper half-plane $\Im z >0$ for all 
contours (the final result is certainly independent of this choice). 
If there are other bulk fields in the correlator, we split the closed 
contour into a small circle $C$ around $z_\d$ and 
a part surrounding the location of all other fields. 
The latter correlation function vanishes 
separately for $\d \ra 0$ due to the ``old'' gluing conditions, whereas 
the former part yields the equation  
\ba 0  & = & \lim_{\d \ra 0}\
             \sum_{n=0}^{\infty} \frac{\lambda^n}{n!} \int_C 
             \cdots \int_C \!\frac{dx_1}{2\pi} \cdots\frac{dx_n}{2\pi}\; 
           \sJ(x_1) \cdots   \sJ(x_n) \nn \\[2mm] 
   & & \hspace*{3cm}  \left( \Phi(w\o \vac;z_\d,\bar z_\d) - 
             \Phi(\vac \o \Omega w; z_\d,\bar z_\d) \right)\ \ .
\nn \ea
Here, we have described the fields $W$ and $\Omega\bW$ in terms of 
the corresponding states $w, \Omega w \in \cV^0$. 
Now we insert the formula (\ref{WOPE}) for 
the operator product expansion between $\sJ$ and the chiral fields. 
Only the residues survive the contour integration so that we get 
$$  \int_C \frac{dx}{2\pi}\; \sJ(x) \bigl( \Phi(w \o \vac;z_\d,\bar z_\d) 
    +  \Phi(\vac \o \Omega w; z_\d,\bar z_\d) \bigr)
     \ = \ i\,\Phi(J_0 w \o \vac;z_\d,\bar z_\d) \ \ . $$
The second term associated with $\Omega \bW$ cannot contribute 
since it is holomorphic in the upper half-plane. Iteration leads to 
\ba  0 & = &  \lim_{\d \ra 0}\ 
             \sum_{n=0}^{\infty} \frac{(i \lambda)^n}{n !} 
              \left( \Phi(J_0^n w \o \vac; z_\d,\bar z_\d ) - 
              \Phi(\vac \o \Omega w;\bar z_\d) \right) \nn \\[2mm]
      & = &  \sum_n\; \bigl(\Phi(\exp(i \lambda J_0) w; z\bigr)
             - \Phi(\Omega w; z) \nn \\[2mm] 
      & =  & \ e^{i \lambda J_0} 
             \, W(z) e^{-i \lambda J_0} - \Omega \bW(\bar z) \ \ .
             \nn
\ea 
Our last step follows from $J_0 \vac = 0$ and the state field 
correspondence for boundary fields. Conjugation with $\exp(i 
\lambda J_0)$ induces an inner automorphism $\gamma_J$ of the chiral 
algebra $\cW$, defined by 
$$ \gamma_J (W) \ := \ \exp(- i \lambda J_0)\,  
 W\, \exp( i \lambda J_0)\ \ \ \mbox{ for all } \ \ \ W \in \cW\ \ .
$$
Replacing $W$ by $\gamma_J (W)$ in the last line of our short 
computation, the final result for the change of the gluing 
conditions under chiral marginal deformations becomes        
\be 
  W(z) \ = \ \Omega \circ \gamma_{\bJ} (\bW) (\bar z) \ \ \ 
  \mbox{ for } \ \ \  z = \bar z \ \ . 
\label{form1} 
\ee
\newline
Observe that $\gamma_J$ acts trivially on the Virasoro field
because a current zero mode $J_0$ commutes with all the modes $L_n$. 
Hence, the gluing condition $T = \bT$ and those 
of all other generators $W \in \cW$ that commute with $J_0$ 
remain unchanged under the deformation with $\sJ$. These fields 
then generate the same Ward identities as before the perturbation. 

\subsubsection{Deformation of the 1-point functions.}
We will now analyze the change of 1-point functions under 
the deformation induced by $\sJ$. Our aim is to derive an 
exact formula for the perturbed 1-point function. 
To this end, we evaluate the terms in eq.\ (\ref{analdef})  order 
by order in $\lambda$ using the operator product expansion 
(\ref{JOPE}) between the field $\varphi_{ij}$ 
and our current $\sJ(x)$. Thereby, the calculation of the 
perturbed 1-point function is essentially reduced to the 
following simple computation:  
\ba \lefteqn{ \langle \varphi_{ij}(z,\bar z) S_J \rangle_\a  =  
    \int_{-\infty}^{\infty} \frac{dx}{2\pi} \, \left(\ \frac{X^i_J}{x-z} \, 
     \langle \varphi_{ij}(z,\bar z)\rangle_\a  - \langle 
     \varphi_{ij}(z,\bar z)\rangle_\a \, 
     \frac{X_{\Omega \bJ}^j}{x-\bar z } \right)} \phantom{XXXXX}\nn \\[2mm]
& = & \int_{-\infty}^{\infty} \frac{dx}{2\pi}  \, \left(\frac{X^i_J}{x-z} 
      \frac{A^\a_{ij}}{(z-\bar z)^{2h_i}}   
     - \frac{A^\a_{ij}}{(z-\bar z)^{2h_i}}  
     \frac{X_{\Omega \bJ}^j}{x-\bar z} \right)\nn \\[2mm] 
& = & \frac{X_J^i A^\a_{ij}}{(z-\bar z)^{2h_i}}
      \int_{-\infty}^{\infty} \frac{dx}{2\pi} \left(
       \frac{1}{x-z} - \frac{1}{x-\bar z}\right)
\  = \  \frac{i\,X_J^i A^\a_{ij}}{(z-\bar z)^{2h_i}} \ \ . \nn 
\ea
\newline
It involves the same kind of arguments as in the previous 
subsection and, in addition, the intertwining relation after  
eq.\ (\ref{1ptfct}). The higher order terms can be computed 
in the same way and give  
\be \langle \varphi_{ij} (z,\bar z) \rangle_{\a:\,\lambda J} 
 \ = \ \frac{e^{i \lambda X_J^i} A^\a_{ij}}{(z-\bar z)^{2h_i}}\ \ . 
\label{form2}
\ee
Consequently, the effect of the perturbation is to ``rotate'' the matrix
$A^\a_{ij}$ with  $\exp(i \lambda X_J^i)$. This behavior 
is consistent with the change of the gluing automorphism and the 
intertwining relation of the linear map 
$A^\a_{ij}$.

\subsubsection{Partition function and cluster property.}
We have argued in the first subsection that correlation functions
involving boundary fields can be deformed by the simple prescription 
(\ref{banaldef}) if all boundary fields in the correlator are local 
with respect to the perturbing field. In the case of a chiral 
marginal perturbation, all boundary fields have this property 
so that formula (\ref{banaldef}) can be used without any 
restriction on the fields $\psi_i$. For correlation functions
without insertions of bulk fields, there are no singularities
in the upper half-plane. Consequently, the effect of the 
deformation on pure boundary correlators is trivial. In 
particular, the conformal weight of all boundary fields is 
unaffected by the perturbation. Hence, {\em the partition function 
$Z_{\Omega,\a)}(q)$ is invariant under chiral deformations}. 
\medskip

Let us also briefly discuss the fate of the cluster property 
(\ref{cluster}) under chiral deformations. Without loss of 
generality, we can restrict ourselves to the investigation 
of 2-point functions.
The basic idea is simple: After expanding the perturbing 
operator $P \exp(S_{\la J})$, we deform the integration 
contours (which originally are parallel to the real axis) 
so that they surround the two insertion points $z_1$ and
$z_2$ in the upper half-plane (see Figure 3). Thereby we 
rewrite the deformed correlation function in each order of 
the perturbation expansion as a sum of unperturbed 2-point 
functions involving descendants of the original bulk fields.
These functions can be split into products of perturbed 
1-point functions by the cluster property of the undeformed 
theory. This last step involves a standard re-summation, 
and the details are left to the reader.  

\begin{figure} 
{\epsfbox{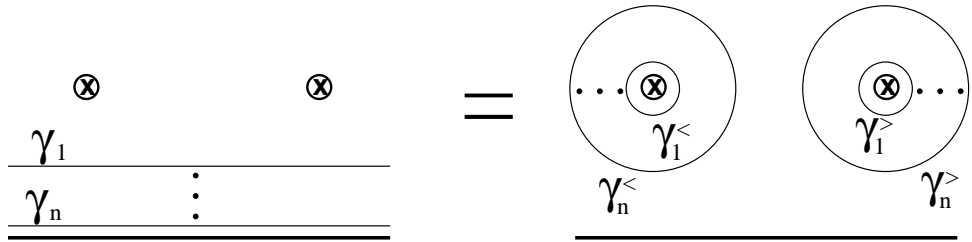}} \vspace*{3mm}  
\caption{ \small For chiral deformations the original curves 
$\gamma_p$ in the contour integrals (\ref{analdef}) can be deformed 
into small circles surrounding the insertion points of two 
bulk fields. The result is expressible through descendants 
of the original bulk fields. } 
\end{figure}

We will see in the next subsection that  our assertions
on chiral deformations can be derived rather easily in the 
boundary state formalism. Here we have chosen an alternative 
(and certainly more cumbersome) route because it allows for 
a first illustration of the prescriptions that underlie 
analytic marginal perturbations. We shall return to more 
general cases in the last subsection after a brief interlude
on the boundary state formalism, which is very effective for 
chiral marginal deformations but difficult to adapt to other cases.

\subsection{The boundary state formalism.}

Most aspects of CFTs on the upper half-plane can be studied equally 
well by introducing boundary states into the ``parent'' CFT on the full 
plane -- more precisely, on the annulus or on the complement of the 
unit disk. Boundary states can be viewed as special linear combinations 
of generalized coherent states (the so-called Ishibashi states), which 
are placed at the boundary of the annulus resp.\ disk complement and which 
provide sources for the bulk fields. This leads to a generalized notion of 
D-branes coupling to closed string modes \cite{ReSc}. 

An abstract characterization of a boundary state can be given in 
terms of bulk field correlation functions: Let us use 
$z, \bar z$ as coordinates on the upper half-plane as before 
and $\xi, \bar\xi$ with 
\be   
\xi \ =\  e^{{{2 \pi i}\over {\beta_0}} \ln z } \ \ \hbox{\rm  and} 
   \ \ \     
 \bar \xi \ =\  e^{-{{ 2 \pi i}\over{\beta_0}} \ln \bar z } 
\label{coortrsf}\ee
to denote coordinates on the annulus within the full complex plane; 
$\beta_0$ is an inverse  ``temperature'', i.e.\ we have identified 
the semi-circles $|z|=t_0$ 
and $|z|=t_0 \exp \beta_0$ with some positive imaginary time $t_0$. 
\footnote{The transformation is easier visualized when split up into 
the two consecutive maps $z \longmapsto w := \ln z$ from the upper 
half-plane to the strip -- hereby the boundary is broken up into 
two components -- and $w \longmapsto \xi$ from the strip to the annulus.} 
\smallskip

The boundary state $|\alpha\rangle$ which implements a boundary condition 
$\alpha$ of the boundary CFT (with Hamiltonian $H^{(H)}\,$) 
into the plane theory (with Hamiltonian $H^{(P)}\,$) is  
defined by demanding the relation \cite{ReSc}
\ba
&\; {\rm Tr}_{{\cal H}_\alpha}\bigl( e^{-\beta_0 H^{(H)}} 
    \varphi^{(H)}_1(z_1,\bar z_1) \cdots 
  \varphi^{(H)}_N(z_N,\bar z_N)\bigr)
  \phantom{XXXXXXXXXXXXXX} \nn \\[2mm] 
&=\ {\cal J}(z,\bar z; \xi, \bar\xi) \cdot  
\langle \Theta \alpha |\,e^{- {{2\pi^2} \over {\beta_0}} \, H^{(P)}}  
     \varphi^{(P)}_1(\xi_1,\bar \xi_1) \cdots 
     \varphi^{(P)}_N(\xi_N, \bar\xi_N)\, | \alpha 
  \rangle\label{corrrel}
\ea
for arbitrary bulk fields $\varphi^{(H)}_i(z_i,\bar z_i) = \varphi_i
(z_i,\bar z_i)$ of the half-plane theory. The Jacobians that appear 
due to the conformal transformation from $(z_i,\bz_i)$ to $(\xi_i, 
\bar \xi_i)$ are collectively denoted by ${\cal J}(z,\bar z; \xi, 
\bar\xi)$; $\Theta$ is the CPT-operator.
The above definition may be extended so as to allow for two different 
boundary states $|\alpha\rangle$, $|\beta\rangle$ at 
the boundaries of the annulus, corresponding to a strip with two 
different boundary conditions $\a,\b$, or to a jump in the 
boundary condition along the real line. 

There exists an alternative way to introduce boundary 
states, namely by equating zero-temperature correlators on the half-plane and 
on the complement of the unit disk in the plane. 
Since this is useful to compute the variation of 1-point functions under 
chiral marginal deformations, we present the formulas. With $z, \bar z$ as 
before, we introduce coordinates $\zeta, \bar\zeta$ on the complement of the 
unit disk by 
\be
\zeta = {1-iz \over 1+i z}\quad\quad {\rm and}  \quad\quad
\bar\zeta = {1+i\bar z \over 1-i \bar z}\ ;
\label{coortrsfzeta}\ee
if $\,|0\rangle$ denotes the vacuum of the bulk CFT, then the requirement 
\ba
&\; \langle\,
    \varphi^{(H)}_1(z_1,\bar z_1) \cdots \varphi^{(H)}_N(z_N,\bar z_N)\,
\rangle_{\alpha}
 \phantom{XXXXxxXXXXXXXxxxxxxX}\nn \\[2mm] 
&=\ {\cal J}(z,\bar z; \zeta, \bar\zeta) \cdot  
\langle 0|\, \varphi^{(P)}_1(\zeta_1,\bar\zeta_1) \cdots 
     \varphi^{(P)}_N(\zeta_N, \bar\zeta_N)\, | \alpha 
\rangle\label{corrrelzeta}
\ea
defines the same boundary states as before;  see e.g.\ \cite{CaLe,ReSc}.

The concrete construction of boundary states 
proceeds in two steps: Given a gluing automorphism $\Omega$ of the chiral 
algebra $\cW$, one first associates Ishibashi states 
$|i\rangle\!\rangle_{\Omega}$ to each pair $(i,\omega^{-1}(i^+))$ of irreducibles 
that occur in the bulk Hilbert space \cite{Ish1}; 
$|i\rangle\!\rangle_{\Omega}$ is unique 
up to a scalar factor (fixed by relation (\ref{ishchar}) below) 
and implements the gluing map in the sense that 
\be
\bigl[\,W_n-(-1)^{h_W}\Omega\bW_{-n}\,\bigr]\,|i\rangle\!\rangle_{\Omega}=0\ .
\label{gluconplane}\ee
Full boundary states $|\alpha\rangle_{\Omega} \equiv |(\Omega,\alpha)\rangle$ 
are given as certain linear combinations of Ishibashi states, 
$$
 |\alpha\rangle_{\Omega} = \sum_i B^i_{\alpha}\, |i\rangle\!\rangle_{\Omega}\ .
$$ 
The complex coefficients $B^i_{\alpha}$ are subject 
to various consistency conditions, most notably to ``Cardy's conditions'' 
arising from world-sheet duality -- see \cite{Car3} for details: The 
partition function of the boundary theory on a strip can be calculated on the 
annulus as a transition amplitude between two boundary states, 
\be
Z_{\alpha \beta}(q) \equiv 
{\rm tr}_{{\cal H}^{(H)}}\bigl( q^{L_0^{(H)}-\frac c {24}} \bigr) 
= \langle \Theta\beta|\,  \tilde q^{L_0^{(P)}-\frac c {24}} \,|\alpha\rangle\ .
\label{partfct}
\ee
This is the two-boundary-state generalization of (\ref{corrrel}) 
without bulk insertions, and  $q=\exp(2\pi i\tau) = \exp(-\beta_0)$, 
$\tilde q = \exp(-2\pi i /\tau)$. The rhs.\ of  eq.\ (\ref{partfct}) can 
be calculated with the help of 
\be
 {}_{\Omega}\langle\!\langle j |\,  \tilde q^{L_0^{(P)}-\frac{c}{24}} 
\,|i\rangle\!\rangle_{\Omega} \ = \ \delta_{i,j}\, \chi^{\cal W}_i(\tilde q)
\label{ishchar}\ee
and, on general grounds,  the lhs.\ in the expression (\ref{partfct}) must 
be a sum of $\cW$-characters with (positive) {\em integer} coefficients. 
After a modular transformation, this implies Cardy's 
non-linear constraints  \cite{Car3} on the coefficients $ B^i_{\alpha}$. In 
particular, the boundary ``states'' should be regarded 
as labels for sectors, not as elements in some vector space. 

With the help of (\ref{corrrelzeta}), one can show \cite{CaLe,ReSc} that 
there is a simple relation to the 1-point functions and structure constants 
of the bulk-boundary OPE -- which are subject to further non-linear sewing 
constraints like (\ref{class}) -- 
namely 
\be
A_{i,\omega^{-1}(i^+)}^{\alpha} = B^{i^+}_{\alpha} \quad\ \ \mbox{and}\ \quad 
C^{\,\alpha}_{\varphi\; 0} 
\ =\ \frac{B^{i^+}_{\alpha}}{B^0_{\alpha}} \ .  
\label{AeqBB}\ee
The decomposition of a boundary state into  Ishibashi states contains 
the same information as  the set of 1-point functions and therefore 
specifies the ``descendant'' boundary CFT of a given bulk CFT completely. 

\medskip

Now let us  exploit the boundary state formalism 
for the discussion of marginal boundary perturbations by 
$W$-algebra currents $J(x)$. To this end, we use eqs.\ (\ref{corrrel}) 
or (\ref{corrrelzeta}) to transport the perturbation from the 
boundary of the upper half-plane to the boundary of the annulus resp.\ 
unit disk. This is possible since $J$ is a local field of the bulk theory 
so that its image under the conformal transformation acts on the state 
space of the bulk theory. With (\ref{coortrsfzeta}) and $h_{J}=1$, we obtain 
\be
\int_{\R} \frac{dx}{2\pi}\, J^{(H)}(x)
= \int_{|\zeta|=1}\! \frac{d\zeta}{2\pi}\,J^{(P)}(\zeta) = i J^{(P)}_0\ .
\label{pertint}\ee
An analogous formula 
results from the map (\ref{coortrsf}) to the annulus, this time the rhs.\ 
consists of one integral  for each boundary component 
at $|\xi|=1$ resp.\ $|\xi|= \frac{2\pi^2}{\beta_0}\,$. 
Since chiral currents are analytic, we need not worry about possible 
divergences, as they can be avoided by deforming the integration contour. 

It is the last equality in (\ref{pertint}) that makes it easy to treat 
perturbations by chiral currents in the boundary state formalism: We 
could {\em not} conclude that $\int_{\R}\!dx\, J^{(H)}(x) = J^{(H)}_0$ 
on the half-plane because of the different integration contour in the 
definition of half-plane modes, see \cite{Car1,Car2,ReSc}. 
Using boundary states, however, the 
effect of marginal boundary perturbations on a half-plane theory 
reduces to the action of current zero modes -- as long as the 
perturbing fields are taken from the chiral algebra. 
\medskip

The boundary states which describe the boundary conditions before 
and after the chiral perturbation are related by a simple ``rotation''. 
Correlators of the deformed boundary CFT can be obtained upon 
replacing $|\alpha\rangle$ in the correspondences (\ref{corrrel}) 
or (\ref{corrrelzeta}) by 
\be 
|(\Omega,\alpha)\rangle_{\lambda\,J} \equiv
|(\Omega,\alpha)\,;\,{\lambda\,J}\rangle \ \;
= \ \,e^{i\lambda J_0}\, |(\Omega,\alpha)\rangle\ 
\label{lambdst}\ee
where $J_0 \equiv  J^{(P)}_0$ is the zero mode of the left-moving current 
on the plane. 

We have made the unperturbed gluing map $\Omega$ explicit in (\ref{lambdst}). 
Indeed, from this formula,  we can immediately re-derive the change 
(\ref{form1}) of the gluing conditions under the marginal deformation by 
$J(x)$: Using that left- and right-movers commute, as well as the simple  
relation $J_n\,|(\Omega,\alpha)\rangle = -\,\Omega\bJ_{-n}\,|(\Omega,
\alpha)\rangle$, eq.\ (\ref{gluconplane}) gets replaced by \vspace*{2mm}
\be
\bigl[\,W_n-(-1)^{h_W}\,\Omega \circ \gamma_{\bJ}\,\bigl(\bW_{-n}\bigr)
\,\bigr]\, |(\Omega,\alpha)\rangle_{\lambda\,J} \ = \ 0\ 
\label{defgluconplane} \vspace*{2mm} 
\ee
with $\gamma_{\bJ}(\bW) := \exp(-i\lambda \bJ_0)\, \bW\, \exp(i\lambda 
\bJ_0)$; see also \cite{GrGu1} for special cases of (\ref{defgluconplane}). 
Likewise, eq.\ (\ref{form2}) for the change of the 1-point functions 
follows from (\ref{corrrelzeta}), (\ref{AeqBB}) and (\ref{lambdst}).
\medskip

Finally, let us use the boundary state formalism to verify -- without 
resorting to perturbative arguments -- that the partition 
function $Z_{\alpha}(q)\equiv Z_{\alpha\alpha}(q)$ of a boundary CFT 
with boundary condition $\alpha$ along the real line stays invariant under 
marginal perturbations by a chiral boundary field $J(x)$: We have to compute
the transition amplitude between $|\alpha;\, \lambda\,J \rangle$ and
$\langle\Theta\,(\alpha;\, \lambda\,J) \,|$. 
But this equals the unperturbed amplitude $Z_{\alpha\alpha}(q)$ because 
$\langle\Theta\, \exp(i\lambda J_0)\,\alpha|= 
\langle\Theta\alpha|\, \exp(-i\lambda J_0)$ and because 
$\exp(i\lambda J_0)$ commutes with $L^{(P)}_0$. 
The spectrum of the boundary theory does not change. 
\smallskip

Up to now, we have always started from a boundary CFT with a constant 
boundary condition $\alpha$ along the real line and considered 
boundary perturbations involving marginal fields that were integrated 
over the whole boundary -- which corresponds to simultaneous deformation 
of one and the same boundary state $|\alpha\rangle$ on both ends of the 
annulus. Generalizations of this would involve jumps in the boundary 
conditions along the real line and different boundary operators 
integrated over the segments of constant 
boundary condition. 

The boundary state formalism allows to discuss the basic case 
with one such jump, using different perturbations for (possibly different) 
in- and out-boundary states in equations (\ref{corrrel},\ref{lambdst}). 
Generically, the partition functions $Z_{(\alpha;\lambda_1J^1),(\beta;
\lambda_2J^2)}(q)$ for such systems will show a different spectrum than 
in the unperturbed situation, and they will involve ``twisted characters'' 
of the symmetry algebra -- more precisely, characters of representations 
twisted by inner automorphisms Ad$_U$ with $U=\exp\{i(\lambda_1J^1_0-
\lambda_2J^2_0)\}$. We shall take advantage of this fact in Section 4.

\subsection{Non-chiral analytic perturbations.} 

Let us now turn towards deformations generated by marginal boundary 
fields $\psi(x)$ that are self-local (in the sense of 
Section 2.4) but {\em not} taken from the chiral algebra. We have 
seen already that these fields are truly marginal to all orders in 
$\lambda$, so we can ask how gluing conditions and 1-point functions 
behave under finite perturbations. We will settle the former issue 
completely in 3.5.1 and make some general statements on 1-point-functions 
and on the spectrum in Subsection 3.5.2.

\subsubsection{Change of the gluing map.}  
As in the case of chiral boundary perturbations, we would like to study 
the effect of non-chiral marginal deformations on the gluing conditions 
$W = \Omega (\bW)$ for the generators of the observable algebra 
${\cal W}$. We start the discussion by showing that $T = \bT$ is not 
changed under analytic deformations to all orders of $\lambda$.  

This follows essentially from the OPE between $\sT(z)$ and $\psi(x)$: 
For a field $\psi$ of conformal dimension $h=1$, the singular part of 
the OPE is a total derivative, 
$$ \sT(z) \, \psi(x) \ = \ \frac{1}{(z-x)^2} \, \psi(x) 
       + \frac{1}{z-x} \, \partial_x \psi(x)\ +\;{\rm reg} \ = \ 
      \partial_x\ \left( \frac{1}{z-x} \, \psi(x) \right)\ +\;{\rm reg} \ \ .
  \ \ 
$$   
We can test the gluing condition for the Virasoro field by inserting
$T(z)$ into the correlation function (\ref{analdef}) such that $\Im z > 
\ve$,  and  then moving $T(z)$  down towards the real axis, where it can 
be compared to $\bT(\bz)$. While passing through one of the contours 
$\gamma_i$, we pick up a term 
$$ \int_C \frac{dx}{2\pi}\ \sT(z) \, \psi(x)\ = \ \int_C \frac{dx}{2\pi}\ 
    \partial_x\, \left( \frac{1}{z-x} \, 
     \psi(x) \right) \ = \ 0 
$$ 
where $C$ is a small circle surrounding the insertion point of the 
Virasoro field. The contour integral along $C$ vanishes, which means 
that the Virasoro field $T$ cannot feel the presence of the 
perturbation and hence the gluing condition stays intact.      
\medskip

The previous argument can be generalized to the following 
simple criterion: 
\newline
{\em Under analytic deformation with a self-local perturbing field 
$\psi(x)$, a prescribed gluing condition for a chiral
field $W(z)$ stays invariant to all orders in $\lambda$ 
if the singular part of the OPE $\;W(z)\,\psi(x)\,$ 
is a total derivative with respect to $x$}. 
\newline
We will encounter several examples later in the text. Let us 
remark that the same criterion is at least necessary for 
other (non-analytic) marginal perturbations to preserve a 
given gluing condition. 

Perturbations with currents $\sJ$ from the chiral algebra often 
lead to a non-trivial deformation (\ref{defgluconplane},\ref{form1}) 
of the gluing condition of a symmetry generator $W(z)$,  
without destroying the associated Ward identity. 
We will see that this is impossible for non-chiral analytic 
deformations:  In Subsection 3.3.1, the change in the 
$W(z)$-gluing condition was obtained by moving the chiral 
field $W(z)$ through the stack of integration contours. 
After a bit of combinatorics, the same procedure for non-chiral 
analytic deformations results in 
$$
W(z) \, e^{\lambda\! \int\!{dx\over2\pi}\,\psi(x)} 
= e^{\lambda\! \int\!{dx\over2\pi}\,\psi(x)}\,  
\Bigl[ e^{\lambda \psi}\,W\Bigr](z)    
$$
(to be understood in the limit $z \ra \bar z$) with 
\be
\Bigl[ e^{\lambda \psi}\,W\Bigr](z)   := 
\sum_{n=0}^{\infty} {\lambda^n\over n!}\; 
\oint_{C_1} \!{dx_1\over 2\pi} \cdots \oint_{C_n} \!{dx_n\over 2\pi} \;
W(z)\,\psi(x_1) \cdots \psi(x_n) \ \ . 
\label{exppsiop}\ee
The curves $C_i$ encircle the point $z$ in the upper half-plane 
as in Figure 3. To each order $n$, the integrals will pick some 
term $\psi^{(n)}$ from the OPE of $W(z)$ with the product of 
perturbing fields. 
At least part of the $\psi^{(n)}$ are true boundary fields which 
are not defined away from the boundary, thus they do not belong 
to the chiral algebra and the above gluing does not produce 
Ward identities for $W(z)$ in the deformed boundary CFT. 

This shows that a non-chiral analytic perturbation either 
breaks or leaves invariant the Ward identity associated to 
a given generator of $\cW$. In general, this leads to a new 
conformally invariant boundary theory with Ward identities 
governed by a subalgebra $\cU$ of the original chiral 
algebra $\cW$. 

\medskip

Let us add a few comments on deformations of boundary 
conditions for $N=2$ superconformal CFTs because they 
constitute an important motivation for the present work
and because they nicely illustrate the criterion given 
above. 
In such theories, one considers two types (A,B) of gluing 
conditions for the chiral fields $G^\pm,J,T$, 
\def\bG{\overline G} 
\ba 
\mbox{A-type:}  & & J(z) \ = \ - \bJ (\bz) \ \ , \ \ 
   G^\pm(z) \ = \eta \, \bG{}^\mp(\bz) \label{Atype}   
  \\[2mm]
\mbox{B-type:}  & & J(z) \ = \ \ \bJ (\bz)\ \ , \ \ 
   G^\pm(z) \ = \ \eta \, \bG{}^\pm(\bz) \label{Btype} 
\ea      
supplemented by $T = \bT$ in both cases. The 
parameter $\eta$ is restricted in order to have a 
supersymmetric ``space-time'' theory. 
More precisely, one requires that an $N=1$ subalgebra 
with generating supercurrent $G(z) := G^+(z) 
+ G^-(z)$ or $G'(z) := i(G^+(z) - G^-(z))$ is preserved 
by any boundary condition. This leaves us with the choice 
$\eta = \pm 1$.  The gluing conditions (\ref{Atype},\ref{Btype}) 
were first introduced in \cite{OOY}, where the connection 
with supersymmetric cycles in Calabi-Yau manifolds was investigated. 
A quite non-trivial realization in CFTs associated with homogeneous 
spaces was constructed in \cite{Stan}. 
\smallskip

It is natural to try and deform an $N=2$ superconformal boundary 
CFT with the chiral U(1) current. According to our general 
formulas, such deformations lead to
$$  \mbox{A type:} \ \ \ G^\pm(z) \ = \ e^{-i\lambda} \eta \,\bG^\mp(\bz) 
   \ , \ \quad \ 
    \mbox{B type:} \ \ \ G^\pm(z) \ = \ e^{i \lambda} \eta \, \bG^\pm(\bz)
\ .
$$
This however, spoils the condition $\eta = \pm 1$ and hence the 
``space-time'' supersymmetry unless $\lambda$ is a multiple of 
$\pi$. Thus there is no family of supersymmetric boundary CFTs 
generated by perturbing an $N=2$ model by the U(1) boundary 
current $J$. 
\medskip

On the other hand, marginal deformations associated with chiral 
or anti-chiral primaries can exist and preserve $N=2$ supersymmetry. 
A state $|\psi_{c,a}\rangle$ (or the corresponding conformal field) 
in an $N=2$ superconformal field theory is called {\em chiral} resp.\ 
{\em anti-chiral} primary if it satisfies 
$$
G^+_{-{1\over2}}\, |\psi_c\rangle = 0 \quad\quad {\rm resp.} \ \ \quad 
G^-_{-{1\over2}}\, |\psi_a\rangle = 0\ .
$$
It follows that  $|\psi_{c,a}\rangle$ are $N=2$ highest weight 
states with charge and dimension related as $q = 2h$ resp.\ $q = -2h$, 
see \cite{LVW}. 
\smallskip

Suppose there is a chiral primary boundary field $\psi_c(x)$ of 
conformal dimension $1/2$ in an $N=2$ boundary CFT, and set 
$\psi(x) := G^-_{-1/2} \psi_c(x) - G^+_{-1/2} \psi_a(x)$ where 
$\psi_a(x) = \bigl(\psi_c(x)\bigr)^*$ is the anti-chiral conjugate 
of $\psi_c(x)$. Then $\psi(x)$ is anti-selfadjoint, uncharged and 
marginal, and we can study the deformations it induces. 
\smallskip

\def\sG{{\sf G}} \def\sbG{{\sf \bG}} 
Typically, there will be other boundary and bulk operators that are 
non-local wrt.\ $\psi(x)$, so we have to rely on the methods developed 
for non-chiral perturbations. The gluing condition for the Virasoro 
field is preserved because of $h_{\psi}=1$ (see above). Since $\psi(x)$ 
carries no charge, the singular contribution to the operator product
of the current $\sJ(w)$ with $\psi(x)$ vanishes so that the gluing 
condition for the current $J$ is untouched. As for the supercurrents 
$\sG^{\pm}(z)$, we use the state-field correspondence and the $N=2$ 
relations to find 
$$
\sG^-(z) \, \Bigl( G^-_{-{1\over2}} \psi_c\Bigr) (x)  \ \sim \ 0\ \ , \ \ \  
\sG^+(z) \, \Bigl( G^-_{-{1\over2}} \psi_c\Bigr) (x) \ \sim \  
\partial_x\,\Bigl( {2 \, \psi_c(x)\over z-x} \Bigr)\  
$$  
together with the analogous relations for the anti-chiral contribution 
$G^+_{-1/2} \psi_a(x)$ to the perturbing field $\psi$. The first equation 
already holds when $\psi_c(x)$ is any $N=2$ primary, whereas in the 
second it is crucial that $\psi_c(x)$ is chiral. Our general criterion 
shows that deformations with $\psi(x)$ do not affect the prescribed 
$N=2$ gluing conditions -- whether they are of A-type or of B-type -- 
to first order in the perturbation parameter; hence they are invariant 
to all orders if $\psi(x)$ is a self-local marginal field. 
\medskip

Deformations induced by chiral primaries as above could serve as 
a starting point to define topological $N=2$ boundary CFTs. In the 
bulk case \cite{DVVtop,War1}, topological field theories yield families 
of commutative associative rings, parameterized by the perturbation 
parameter, which often can be interpreted as quantum cohomology 
rings of complex manifolds. It would be interesting to see which
new structures arise from topological boundary correlators. Since 
the topology of the ``supporting space'', i.e.\ of the world-sheet 
boundary, does not allow to continuously interchange arguments in 
correlation functions, one may expect that non-commutative rings 
appear quite naturally. 

\subsubsection{One-point-functions, spectrum, and the cluster property.} 
A boundary conformal field theory is determined by the gluing conditions
and the 1-point functions. We have discussed the change of gluing conditions 
under non-chiral analytic deformations, but it is difficult to obtain general 
statements on the deformed 1-point functions, in particular because they 
are to be computed for all primary fields of the smaller (``unbroken'') 
subalgebra $\cU \subset \cW$ associated to the reduced set of Ward 
identities that may survive after turning on the perturbation. 
Nevertheless, as we will see later on, there are  examples of 
non-trivial analytic deformations for which the deformed 1-point 
functions can be constructed to all orders.
 
At the moment, we limit ourselves to a simple first order criterion 
for the {\em invariance} of a 1-point function. Let $\varphi(z,\bar z)$ 
be an arbitrary quasi-primary bulk field, e.g.\ a primary field 
for the reduced chiral algebra $\cU \subset \cW$. Conformal 
transformation properties fix the 2-point function of $\varphi(z,
\bar z)$ with the perturbing field $\psi(x)$ up to 
a constant, 
$$ \langle \varphi(z,\bar z) \psi(x) \rangle_\a \ = \ 
   \frac{C^\a_{\varphi\psi}}{(z-\bar z)^{2h-1} (z-x)(\bar z -x)} 
   \ \ . $$
Here, $h = \bar h$ is the conformal weight of the field $\varphi$,  
and the bulk-boundary OPE coefficient  $C^\a_{\varphi \psi}$ depends 
on the original boundary condition $\a$. By the residue theorem, we 
get the following first order correction for the perturbed 1-point 
function 
\be 
\langle \varphi(z,\bar z) \rangle_{\a;\,\lambda \psi} \ = \ 
\langle \varphi(z,\bar z) \rangle_\a + i\,\lambda\, 
\frac{C^\a_{\varphi\psi}} {(z-\bar z)^{2h}} + O(\lambda^2) \ \ .
\ee
To leading order, a 1-point function $\langle 
\varphi(z,\bar z) \rangle_\a$  is invariant under a 
perturbation with $\psi$ if and only if $C^\a_{\varphi\psi} = 0$. 
Again this is a necessary condition for the invariance of a
given 1-point function under any truly marginal perturbation,  
but it is certainly not sufficient.   
\medskip 

A full computation of the partition function requires complete
knowledge of all 1-point functions and hence it is at best 
accessible through a case by case study. On the other hand, 
there are some general statements we can make about the behaviour
of $Z_\a (q)$ under analytic deformations. We have argued above 
that the formula (\ref{banaldef}) can be used to construct perturbed  
correlators of boundary fields $\psi_i$ which are local with 
respect to the perturbing field $\psi$. By the same arguments 
as in chiral deformation theory, we conclude that the conformal 
weights of such fields $\psi_i$ are invariant under the 
deformation. While this criterion does not protect the 
full spectrum of boundary conformal weights (as in the 
case of chiral deformations where all boundary fields are 
local with respect to $\psi$), it shows that {\em part of 
the partition function stays intact}. In particular, all 
chiral fields $\sW$ are local with respect to $\psi$ so 
that the partition function will always contain the 
vacuum character of the original chiral algebra $\cW$ {\em even 
if}  gluing conditions and Ward identities are broken down to 
a subalgebra $\cU \subset \cW$. Furthermore, while the 
``gluing'' (\ref{exppsiop}) of a chiral field $W(z)$  to boundary 
operators destroys the Ward-identity for $W(z)$, it still leads 
to a (possibly twisted) action of the full chiral algebra $\cW$ 
on the state space $\cH$. This effect can be read off from 
the partition function of the deformed theory which still
decomposes into characters of (twisted) representations
of $\cW$, see the examples below. 
\medskip

The cluster property is somewhat more difficult to attack.
Note that the argument at the end of Subsection 3.3.3 cannot 
be used in this simple form because the deformed correlators 
are not expressible through correlators of descendants of 
the original bulk fields. There exists a variant of the 
previous reasoning which takes into account the specific 
analyticity properties of correlators with insertions 
of self-local non-chiral boundary fields and bulk fields. Its 
convergence behaviour in the limit $n \rightarrow 
\infty$, however, is not easy to control. It is likely 
that the cluster property is preserved for an open 
neighbourhood of $\lambda = 0$ but is bound to break 
down at certain finite values of the perturbation parameter 
$\lambda$ whenever we deform with some non-chiral boundary 
fields. This agrees with the examples we analyse below. 
Often, the breakdown of the cluster property has 
an interesting physical or geometric interpretation.

\section{Example: Boundary deformations for {\fatma c}$\;$=$\;$1 theories} 

The results of the previous section hold for arbitrary boundary CFTs. 
We will now illustrate them in a  simple example, namely the free bosonic 
field. To begin with, we present the uncompactified theory with Neumann 
and Dirichlet boundary conditions and study their deformations. Then the 
same analysis is made for the compactified boson. In the third subsection, 
we investigate boundary perturbations of $c=1$ orbifold theories. 
Although the models under consideration are simple enough, we will 
encounter rich patterns in the brane moduli space, including some 
unexpected phenomena.

\subsection{The uncompactified theory.} 
The dynamical degrees of freedom of the bulk theory 
are obtained from a single field $X(z,\bar z)$ which obeys the 
usual equation of motion $ \partial \bar \partial X(z,\bar z) = 0$. 
The modes of the left- and right-moving chiral currents 
$J(z) = 2 i\, \partial X(z,\bar z) =  \sum \, a_n\, z^{-n-1}$ 
and $\bJ(\bar z) = 2i\, \bar \partial X(z,\bar z)=  
\sum  \baa_n \,\bar z^{-n-1}$
generate a U(1)$\,\times\,$U(1) algebra with canonical 
commutation relations 
$$ 
   [\, a_n\, ,\, a_m\, ] \ = \ n\, \d_{n,-m} \ \ , \ \ 
   [\, \baa_m \, ,\, \baa_m\, ] \ = \ n\, \d_{n,-m}\ \ . 
$$
The Virasoro fields are obtained from $J,\bJ$ by 
normal-ordering, $ T(z) =  \frac12\, {\bf :}\, J\,J\,{\bf :}\,(z)$ 
and $\bT(\bar z)  = \frac12\, {\bf :}\, \bJ\, \bJ\,{\bf :}\,(\bar z)\;$. 
\newline
The abelian current algebra has irreducible representations $\cV^g$
labeled by real numbers $g$, the U(1) charge. $\cV^g$ is 
generated from a ground state $|g\rangle$ with the 
properties 
$$ a_n\;|g\rangle \ = \ 0 \ \ \mbox{ for all } \ \ n > 0 \ \ 
   \ \mbox{ and } \ \ a_0 \; |g\rangle \ = \ g\; |g \rangle \ \ .$$
The lowest-energy subspace $V^g_0$  of $\cV^g$ is one-dimensional 
and spanned by $|g\rangle$, the element $a_0$ acts on $V^g_0$ by 
$X^g_J = g$.   

Putting things together, one can realize  
the bosonic field $X$ on the state space $\cH^{(P)} = \bigoplus_g
\cV^g \o \cV^g$ which is a diagonal sum with equal 
U(1) charges for both chiralities. In the explicit 
formula 
$$ X(z,\bar z) \ = \ x - \frac{i}{4}\,p\, \ln (z\bar z) + 
    \frac{i}{2} \sum_{n \neq 0} \, \left( \frac{a_n}{n} \, z^{-n} + 
    \frac{\baa_n}{n} \, \bar z^{-n} \right)\ , $$
one new element $x$ appears which acts as differentiation 
$x = i \partial_g$ on the state space. We have also introduced 
the operator $p =  a_0 + \baa_0 $ which has the usual 
Heisenberg commutation relation with $x$.  Bulk fields
$\varphi_{g_1g_2}(z, \bar z)$ exist only for $g_1 = g_2$ so that 
we will omit one index in the following. $\varphi_g = 
\varphi_{gg}$ is obtained from the bosonic field by 
$$   \varphi_g(z,\bar z) \ = \ {\bf:}\,\exp(2 i g X(z,\bar z))\,{\bf:} \ \ , 
$$
and with proper normal-ordering these fields can be shown to 
obey the operator product expansions 
$$ \varphi_{g_1}(z_1,\bar z_1) \ \varphi_{g_2}(z_2, \bar z_2) \ 
    \sim \ |z_1 - z_2|^{2 g_1 g_2}  \varphi_{g_1 +g_2} 
    (z_2,\bar z_2) + \dots \ \ . $$
The conformal weights $h_g = \bar h_g$ of $\varphi_g(z,\bar z)$ 
are given by 
$h_g = \frac12 g^2$. 
\bigskip 

We will look for boundary conditions that preserve the chiral 
symmetry algebra generated by the U(1) current $J$. There are 
two possibilities for the gluing map which we can use:
\ba
   \mbox{Neumann boundary condition:} \ \ \ \ \ \  
      J(z) & = &  \Omega_N\bJ(\bar z) \ \equiv \ \ \bJ(\bar z) \
   \label{N} \\[2mm]
   \mbox{Dirichlet \ boundary condition:} \ \ \ \ \ \
      J(z)  & = &  \Omega_D\bJ(\bar z) \ \equiv \ - \bJ(\bar z) \
    \label{D}
\ea
The Neumann type boundary conditions are realized by a bosonic 
field 
$$ X(z,\bar z) \ = \ x -  \frac{i}{4}\,p\, \ln (z\bar z) + 
   \frac{i}{2} \sum_{n\neq 0} 
      \, \frac{a_n}{n}\ \left( \, z^{-n}  + \bar z^{-n}\, \right)  $$ 
acting on a state space $\cH = \bigoplus_g \cV^g$. Here, 
$x = i \partial_g$ as before, and $a_n$, $p= 2 a_0$ are the modes of the 
generator $\sJ$ of the U(1) symmetry in the boundary Hilbert space. 
The computation of the 1-point 
function of $\varphi_g(z,\bar z) =\; {\bf :}\,\exp(2 i g X(z,\bar z))
 \,{\bf :}$ is a straightforward exercise, leading to 
$$ \langle \varphi_g(z,\bar z) \rangle_N \ = \ \delta_{g,0} \ \ .$$ 
Note that there appears no free parameter in these 1-point 
functions, i.e.\ there is only one boundary theory with Neumann 
boundary conditions for an uncompactified free boson. 
\smallskip

For Dirichlet boundary conditions, we build the bosonic 
field according to    
$$ 
    X(z,\bar z) \ = \ x_0 + \frac{i}{2} \sum_{n\neq 0} 
      \,  \frac{a_n}{n}\ \left(  z^{-n} - \bar z^{-n} \right)
     \ \ .  
$$  
Here, $x_0$ is a free real parameter describing the 
value of the bosonic field along the boundary, i.e.\ 
$X(z,\bar z) \equiv X(z)+X(\bar z) = x_0$ for $z = \bar z$. The field 
$X$ acts on a state space $\cH = \cV^0$ consisting only of 
the vacuum representation. This time, calculation of 1-point 
functions for $\varphi_g(z,\bar z)$ results in 
the formula 
$$ \langle \varphi_g(z,\bar z) \rangle_{D\, x_0} 
  \ = \ \frac{e^{2 i g x_0}}{(z-\bar z)^{2h_g}} \ \  
$$ 
which depends on $x_0$, parameterizing different possible 
boundary theories with Dirichlet boundary conditions.           

Note that, for the free boson theory, the structure constants
$\Xi$ in the sewing constraint (\ref{class}) are given by 
$\Xi^{g'}_{g_1,g_2} = 
\delta_{g', g_1+g_2}$. The numbers $A^{x_0}_g
= \exp(2 i g x_0)$ solve eq.\ (\ref{class}) and hence the 
theory has the cluster property (\ref{cluster}); at the same time, 
this means that superpositions (``mixtures'') of ``pure'' Dirichlet 
boundary conditions do not cluster. 
\medskip 

\subsubsection{Chiral deformations.}  
Let us now study marginal deformations and start with 
the chiral current $\sJ$ which is the only field
of weight $h=1$ in the chiral algebra. Since the 
zero mode $a_0$ commutes with all other elements 
in $\cW$, it generates the trivial inner automorphism
$\gamma_J = id$ on the chiral algebra. It follows then 
from eq.\ (\ref{form1}) that the gluing conditions 
are invariant under the 
deformation, i.e.\  they are given by the formulas (\ref{N},\ref{D}) 
for all values of the perturbation parameter 
$\lambda$. Consequently, the only possible
effect of the perturbation on the boundary theories 
is due to changes of the 1-point functions. 

For Neumann boundary conditions we have $A^N_{g} = 
\delta_{g,0}$ so that, according to our formula 
(\ref{form2}), this coefficient -- and therefore
the Neumann boundary theory -- stays invariant under 
deformations with $\sJ(x)$. For Dirichlet boundary 
conditions, things are a bit more interesting. The coefficients 
$A^{x_o}_{g} = e^{2 i g x_0}$ of the Dirichlet boundary 
condition behave as 
$$ A^{x_0}_{g} \ \longrightarrow \ e^{i \lambda g}  
   \, e^{2 i g x_0}
   \ = \ e^{2 i g (x_0+\frac{\lambda}{2})} \ \  $$
when we turn on the perturbation. As a result of 
the deformation, the parameter $x_0$ gets shifted by 
$\lambda/2$ -- i.e.\ the D-brane is displaced. 

\subsubsection{Non-chiral deformations.}
In the case of Neumann boundary conditions, there are two other 
boundary fields of conformal dimension $h=1$. We will consider
perturbations by the combinations 
$$
\psi^1(x) \ := \ \sqrt{2} \,\cos \{2 \sqrt{2} X(x)\} \ \ \ \ \ 
\mbox{ and }\\ \ \ \ 
\psi^2(x) \ :=\  \sqrt{2} \,\sin \{2 \sqrt{2} X(x)\}\ ,
$$
which will be seen to break the chiral symmetry down to 
the Virasoro algebra by inducing a periodic ``potential'' 
along the boundary. This has been studied in some detail in 
\cite{CaKle,CKLM,PoTh}. 

The boundary fields $\psi^a(x)$, $a=1,2$, are local with respect 
to themselves. By our general considerations of Section 3.2 
on analytic perturbations, $\psi^a(x)$ are truly marginal (to all 
orders in the ``coupling'' $\lambda$). At the same time, we 
expect the spectrum of boundary fields to change when the boundary 
potential is turned on, because the boundary Hilbert space $\cH = 
\bigoplus_g {\cal V}^g$ of the Neumann theory contains fields which 
are non-local wrt.\  $\psi^a(x)$ -- in fact, only the scaling 
dimensions of operators with charges in $\sqrt{2}\Z$ are protected.
\medskip

Let us first see how the U(1) gluing conditions 
behave under perturbations with e.g.\ $\psi^1(x)$. The criterion for 
invariance of $\Omega$ given in Section 3.5 required that the 
singular part of the OPE between a chiral symmetry generator $W(z)$ 
and $\psi^1(x)$ is a total $\partial_x$-derivative. This is true for 
the Virasoro field $W(z) = T(z)$, but not for the current $W(z)=J(z)$. 
So we have to determine the effect of pushing $J(z)$ through the 
$x_i$-integration contour when moving the field towards the real 
line, in order to evaluate (\ref{exppsiop}) describing the change 
of  $J(z) = \overline{J}(\bar z)$. The OPE of $J(z)$ with $\psi^1(x)$
is given by 
\be
J(z) \, \psi^1(x) \ = \ { i \sqrt{2} \over z-x}  \,\psi^2(z) \ +\ {\rm reg}
\label{JjOPE}\ee
so that we pick up a term $i \sqrt{2} \psi^2(z)$ whenever $J(z)$ 
passes one of the contours. The effect of moving the field $\psi^2(z)$ 
towards the real axis is determined by the OPE 
\be
\psi^2(z)\, \psi^1(x) \ = \ {-i \sqrt{2}  \over z-x}\, J(z)  +\ {\rm reg}
\ \ . \label{jjOPE}
\ee
We can now apply our general formula (\ref{exppsiop}) to derive 
the following closed expression for the $\psi^1$-deformed gluing 
conditions:
\be
J(z) \ = \ \sin(\sqrt{2} \lambda)\, \psi^2(x) + \cos(\sqrt{2} \lambda)\, 
\overline{J}(\bar z)
\label{newglcond}\ee 
for $z=\bar z = x$. By the same reasoning, one can determine the 
effect of perturbations with $\psi^2$ on the Neumann gluing condition:
\be
J(z) \ = \ - \sin(\sqrt{2} \lambda)\, \psi^1(x) + \cos(\sqrt{2} \lambda) \,
\overline{J}(\bar z)\ .
\label{newglcondtwo}\ee 
These equations, which were also found in \cite{CKLM}, mean that 
the boundary reflects left-moving into right-moving currents 
only at the expense of marginal boundary fields. As a consequence, 
the correlation functions of the perturbed boundary CFT no longer 
obey Ward identities for the U(1) current $\sJ$. Exceptions occur 
whenever $\lambda = n\,{\pi\over\sqrt{2}}$ for some integer $n$: 
Then $\psi^a(x)$  disappear from (\ref{newglcond},\ref{newglcondtwo}) 
and, if $n$ is odd, the original Neumann conditions for $J(z)$ 
are turned into Dirichlet conditions. We will refer to the latter 
values of the perturbation parameter as {\em Dirichlet-like} 
points. 

Broken U(1) symmetry 
complicates computations considerably. Since it is only the 
Virasoro algebra  that remains at our disposal, we have to 
characterize the deformed theory through the 1-point functions
of all Virasoro primary fields. The decomposition of irreducible 
U(1) modules into $c=1$ Virasoro modules is well known.  
Both coincide as long as the 
conformal dimension of the primary field is not given by 
$h = {m^2}$ for any $m \in \frac12 \Z$, but for those cases, 
the U(1) modules decompose into a sequence of irreducible 
Virasoro representations, 
\be
\cV^{\rm U(1)}_{{\sqrt{2} m}} \ = \ \bigoplus_{l=0}^{\infty} 
\cV^{\rm Vir}_{{(|m|+l)^2}}
\label{UVirdecomp}\ee
-- the subscript on U(1) modules is the charge, the one on 
Virasoro modules the conformal dimension. There is   
a corresponding identity for the characters,  
\be\chi^{\rm Vir}_{{m^2}}(q) \ = \ \chi^{\rm 
  U(1)}_{{\sqrt{2}m }}(q) -\chi^{\rm U(1)}_{ \sqrt{2}{(|m|+1)}}(q)\ .
\label{excepVirchars}\ee
It means that, for the values $h={m^2},\ m \in \frac12 \Z$, the 
$c=1$ Virasoro Verma modules contain a singular vector at level 
$2|m|+1$. \smallskip

Coming back to our problem of describing the deformed boundary 
theories, we first remark that the theory has a rather 
useful ``hidden'' SU(2) symmetry which also governs the deformed 
theories. In fact, this symmetry is obvious from the OPEs of $J, 
\psi^1, \psi^2$ which, while not forming an algebra of true local 
currents for the full boundary CFT, still lead to the same 
algebraic structure for various quantities of interest, in 
particular for 1-point functions of bulk fields. This SU(2) 
symmetry is also visible in the structure of the decomposition 
(\ref{UVirdecomp}). Indeed, the Virasoro highest weight vectors at 
energy $h = j^2$ in the state space $\cH$ of the Neumann theory 
span an SU(2) multiplet of length $2j+1$ so that  
$$ \cH \ = \ \int_{g \neq \sqrt{2}m} \cV^{\rm U(1)}_g \ \oplus \ 
    \bigoplus_{m \in \frac12 \Z} \cV^{\rm U(1)}_{\sqrt{2} m }
  \ = \ \int_{g \neq \sqrt{2}m } \cV^{\rm Vir}_{g^2 \over 2}
 \ \oplus \ \bigoplus_{j\in\oh \Z_+}^{\infty} \Bigl(\cV^{\rm Vir}_{{j^2}}  
   \Bigr)^{\oplus\,2 j +1}\ .
$$
A similar structure is observed for the state space $\cH^{(P)}$ 
of the bulk theory, 
\ba
\cH^{(P)}  & = &  \int_{g \neq \sqrt{2}m } \cV^{\rm U(1)}_g \o  \
  \cV^{\rm U(1)}_g \ \oplus \ \bigoplus_{m \in \frac12 \Z} 
\cV^{\rm U(1)}_{{\sqrt{2}m}}\otimes\cV^{\rm U(1)}_{{\sqrt{2}m}}
\nn \\[2mm] 
 & = &  \int_{g \neq \sqrt{2}m} \cV^{Vir}_{g^2 \over 2} \o 
     \cV^{Vir}_{g^2 \over 2} \ \oplus \
\bigoplus_{j\in{\oh}\Z_+} \Bigl(\cV^{\rm Vir}_{{j^2}} 
\otimes \cV^{\rm Vir}_{{j^2}} \Bigr)^{\oplus\,2 j +1} \ 
\oplus \ \ldots
\label{flatbulkVir}
\ea
where the dots denote terms with $h\neq \bar h$, which are of 
no concern to us since they cannot couple to a conformal 
boundary state. {}From these formulas we conclude that spin-less 
(i.e.\ $h = \bar h$) Virasoro primary bulk fields come in two 
families:
$$ (1) \ \ \ \varphi_{g,g} (z,\bar z) \ \ \ \mbox{ with } \ \ \ 
              g \not \in{\textstyle \frac{1}{\sqrt{2}}}\Z \ \ \ \ \ \ \ \ \ 
(2) \ \ \ \varphi^j_{m,m} (z,\bar z) \ \ \mbox{ with } 
              j \in {\textstyle \frac12} \Z_+ 
$$ 
and $m = -j, -j+1, \dots , j-1,j$. The fields of the second 
family have U(1) charges $g = \bar g = \sqrt{2} m \in 
{\textstyle\frac{1}{\sqrt{2}}} \Z$ with respect to $J_0$. 
\smallskip

Since the perturbing fields span the charge lattice 
$\sqrt{2} \Z$, U(1) charge conservation implies that the 
1-point functions of fields $\varphi_{g,g}$ in the first 
family are not perturbed, i.e.\ 
\be 
\langle \varphi_{g,g}(z,\bar z)\rangle_{N;\lambda \psi^a} 
   \ = \ 0 \ \ \  \mbox{ for } \ \ \ 
g \not \in {\textstyle\frac{1}{\sqrt{2}}} \Z \ \ . 
\label{unchonept}\ee
For the fields $\varphi^j_{m,m}$, results get more interesting. 
In the evaluation of the deformed correlators we continue
the perturbing field analytically into the upper half-plane 
and compute the usual contour integrals. This leads to an action 
of the SU(2) generators $J_0, \psi^a_0 := \int\!\frac{dx}{2\pi}\,\psi^a(x)$ 
on the left index of the fields, i.e.\  
\be
\langle \varphi^{\,j}_{m,m}(z,\bar z) 
\rangle_{N;\,\lambda \psi^a} \ =\  
\sum_{m'=-j}^j D^{\,j}_{m,m'} (\Gamma^a_{\lambda}) \;
\langle \varphi^j_{m',m}(z,\bar z) \rangle_{N}
\label{pertonepoint}\ee
where $\Gamma^a_{\lambda} = \exp(i\lambda \psi^a_0)$ is regarded as an 
SU(2)-element, and $D^{\,j}_{m,m'} (\Gamma^a_{\lambda})$ are the 
entries of its spin $j$ representation matrix expressed in a 
spin$_z$ eigenbasis. Finally, the correlator on the rhs.\ of 
(\ref{pertonepoint}) stands for the function  
$$
\langle \varphi^{\,j}_{m',m}(z,\bar z) \rangle_{N} \ = \ 
\delta_{m', -m}\,  {1\over (z-\bar z)^{m^2} }\ ,
$$
even if $\varphi^{\,j}_{-m,m}(z,\bar z)$ does not occur in the 
uncompactified free boson theory. 
We can also encode the outcome of this computation in the 
following formula for the $\psi^a$-perturbed flat Neumann 
boundary state: 
\be
|N;\,\lambda \psi^a\rangle \ = \  \sum_{j\in{\oh}\Z_+}
\sum_{m=-j}^j \, D^{\,j}_{m,-m} (\Gamma^a_{\lambda}) \,|j,m,m \rangle\!
\rangle
\label{flatpertbdst}
\ee
where $ |j,m,m \rangle\!\rangle$ are Virasoro Ishibashi states associated 
to the primaries $ \varphi^{\,j}_{m,m}(z,\bar z)$. 
\medskip

While (\ref{flatpertbdst}) in principle gives complete information on the 
perturbed boundary theory, it looks essentially hopeless to compute 
the perturbed partition function $Z_{\alpha}(q)$ directly via a 
modular transformation of $\sum_{j,m} | D^{\,j}_{m,-m} 
(\Gamma^a_{\lambda}) |^2 \chi^{{\rm Vir}}_{j^2}(\tilde q)$ -- simply 
because the matrix elements $D^{\,j}_{m,n} (\Gamma)$ are given by the
rather cumbersome formula  
\ba
D^{\,j}_{m,n} (\Gamma) & = & \sum_{\mu={\rm max}(0,n-m)}^{{\rm min}
(j-m,j+n)}\  {[(j+m)!(j-m)![(j+n)!(j-n)!]^{\oh} \over
(j-m-\mu)!(j+n-\mu)! \mu! (m-n+\mu)! } \, \phantom{xxxxxxxxxx}
\nn \\[2mm] & & \phantom{xxxxxxxxxxxxxxxx}\times\; 
a^{j+n-\mu}\, (a^*)^{j-m-\mu} \, b^{\mu}\, (-b^*)^{m-n+\mu} \label{Djmn}
\ea
in which the group element $ \Gamma \in$ SU(2) was parameterized 
by $\Gamma = { a\ b \choose -b^* a^* }\,$; see e.g.\ \cite{Ham}. At 
the Dirichlet-like points $\lambda = \,{2k+1\over\sqrt{2}}\pi,\ k\in\Z$, 
however, the formula simplifies considerably, and modular 
transformation yields 
\be
Z_{\alpha_D}(q)\  \sim \ \sum_{n\in\Z} \; {q^{n^2} \over \eta(q) } 
\label{Zdirlike}\ee
for $\alpha_D =$ Dirichlet-like boundary conditions: The initially 
continuous Neumann spectrum is reduced to a discrete one (which 
furthermore is the same as the one of a boundary CFT of a free boson 
compactified at the self-dual radius). The boundary condition can be 
viewed as a superposition of flat D-branes located at the sites of an 
infinite lattice.  The boundary fields with non-zero U(1) charge should 
be attributed to ``solitons'' interpolating between different 
minima of the boundary potential. 
\smallskip

In \cite{PoTh}, the partition function for an arbitrary perturbation was 
computed along a different route, namely by passing to a free fermion 
representation and by explicitly diagonalizing the Hamiltonian consisting 
of a free part and the boundary interaction. Since the technical details 
are not very illuminating, we merely state their result: The spectrum of 
the perturbed Neumann boundary states $|\alpha\rangle_\lambda := 
|N;\,\lambda \psi^1\rangle$ is given by 
\be
Z_{\alpha_\lambda}(q) 
\ = \ \eta(q)^{-1}\;\sum_{m\in\Z} \int_{0}^2 \!d\zeta \;
  q^{(2m +f_{\lambda}(\zeta))^2} 
\label{ZPolTh}\ee
with 
\be
f_{\lambda}(\zeta) \ = \ {1\over \pi} 
\arcsin[\, \cos{{\textstyle \frac{\lambda}{\sqrt{2}}}} \,\cdot\, \sin\pi \zeta \,]\ ;
\label{arcsinfunc}
\ee
the arcsin-branch is to be chosen such that 
$\lim_{\lambda\to0} f_\lambda(\zeta) = \zeta$; the integral is over the 
half-open interval, which becomes important in the discrete variants to be 
discussed below. (\ref{ZPolTh}) displays a band structure of the 
spectrum which is typical for a theory of electrons moving 
in a crystal. As soon as an infinitesimal periodic potential is 
turned on, the continuous spectrum rips apart at the values 
$h= \frac{n^2}4$,  and the gaps open up as the strength $\lambda$ of the 
potential grows. The bands are reduced to points at the Dirichlet-like value 
$\lambda={\pi\over\sqrt{2}}$, where only primaries with dimension 
$h=n^2$ for $n \in \Z$ remain (tight-binding limit).
Naively one would expect this to occur at $\lambda = \infty$ 
but, loosely speaking, the period of the potential introduces a 
``scale'' $r_{\rm s.d.} = \frac{1}{\sqrt{2}}$ into the problem 
so that special effects are bound to appear whenever $\lambda$ is 
in resonance with $r_{\rm s.d.}$.  
\smallskip

The structure of the spectrum is in line with our general expectation. 
In fact, it does decompose into U(1) characters and all states of 
U(1) charge in the lattice $\sqrt{2} \Z$ -- which correspond to 
fields that are local with respect to the perturbing fields -- 
do remain in the boundary theory.
\medskip

The physical interpretation of the periodic boundary potential with 
Dirichlet-like coupling strength as generating a mixture 
of elementary Dirichlet conditions is rather compelling, 
but suggests that the perturbed boundary theory violates the 
cluster property. Indeed, at  $\lambda={2k+1\over\sqrt{2}}\pi$, 
the cluster relation together with the Dirichlet Ward identities 
for the U(1) current would imply the sewing constraint   
\be
A^{\alpha_{D}}_{g_1} \, A^{\alpha_{D}}_{g_2}
  \ \ {\buildrel ? \over =} \ \ A^{\alpha_{D}}_{g_1+g_2}\ .
\label{clcondflat}\ee
Choosing $g_1,g_2$ such that $g_i \notin \frac{1}{\sqrt{2}} \Z$ 
but $g_1+g_2 = \frac{1}{\sqrt{2}}$, the structure constants 
$A^{\a_D}_{g_i}$ vanish as in the original Neumann boundary theory, 
cf.\ (\ref{unchonept}), while  $A^{\a_D}_{g_1+g_2}\neq 0$. 

In order to test clustering for arbitrary values of 
$\lambda$, we would need a lot of information on fusion and chiral blocks 
of $c=1$ Virasoro modules, about which virtually nothing is known. 
We expect, however, that the boundary states (\ref{flatpertbdst}) 
obey the cluster condition as long as $|\lambda| < \frac{\pi}{\sqrt{2}}$, 
for the following reasons: Our study of orbi\-fold models will show that 
this is true for the $r_{\rm circ}= \sqrt{2}$ circle model  -- 
which possesses analogous deformations with exactly the same algebraic 
properties as the $r=\infty$ theory. Furthermore, the general argument in 
favour of clustering which was sketched in Subsections 3.3.3 and 3.5.3 
indicates that finite domains of convergence could spoil clustering at 
finite perturbation strength.

\subsection{The compactified theory.}
If we take a circle of radius $r$ as the target space for 
the free boson, we can again impose Dirichlet and Neumann 
boundary conditions, but now there are continuous parameters 
in both cases. Compared to the uncompactified case, the mode 
expansion of the bulk field $X(z,\bar z)= X_L(z) + X_R(\bar z)$ 
additionally involves a winding number operator $w$ as well as two 
independent zero mode operators $x_{L,R}$: 
\ba 
 X_L(z) & = & x_L - \frac i 4\ p\;  \ln z - \frac i 2\  r\,w \; 
\ln z + \frac i 2 \     \sum_{n \neq 0} \, \frac{a_n}{n} \, z^{-n} 
\ \ , \ \ \nn \\[2mm]   
 X_R(\bar z) & = & x_R - \frac i 4\ p\;  \ln \bar z + \frac i 2\  r\,w \; 
\ln \bar z + \frac i 2 \  \sum_{n \neq 0} \, \frac{\baa_n}{n} \, \bar z^{-n}  
\ \ . \nn 
\ea 
The normalizations are chosen so as to preserve the canonical commutation 
relations from the uncompactified case: The winding operator $w$ commutes 
with $x:=x_L+x_R$, $p$ and all oscillators, all other relations follow 
from the exchanges $p \leftrightarrow 2rw\,,\ x \leftrightarrow \tilde 
x:= x_L-x_R$. For later convenience, we introduce the zero modes $a_0 
:=  p/2 + r w$ and $\overline{a}_0 := p/2 - r w$. Chiral currents 
$J(z)$ and $\overline{J}(\bar z)$ are obtained from the modes $a_n, 
\baa_n$ as before. 

Because of the new degree of freedom, primary fields 
$\varphi_{g,\bar g}(z,\bar z) = e^{2ig X_L(z)} e^{2i\bar gX_R(\bar z)}$ 
can carry different left- and right-moving charges wrt.\  $a_0$ and 
$\overline{a}_0$, namely $g=k/2r + rw$ and $\bar g =k/2r - rw$, where 
$k :=rp$ and $w$ take integer values. Again one can easily solve the 
Dirichlet and Neumann boundary conditions eqs.\ (\ref{D}) and (\ref{N}) 
in terms of the bosonic field with values on the circle to arrive 
at  
$$
 \langle \varphi_{g,\bar g}(z,\bar z) \rangle_{D\, x_0} 
  \ = \ \delta_{g,\bar g}\; \frac 1 {\sqrt{2r}} \; \frac {e^{i k x_0/r}} 
{(z-\bar z)^{ \frac{k^2}{4r^2}} } \ \  
$$ 
for the Dirichlet case -- the real parameter $x_0 \in \R\; {\rm mod}\; 2\pi r$ 
can again be interpreted as the location of the brane, i.e.\ $X(z) = x_0 - 
X(\bar z)$ for $z= \bar z$. The Neumann case is obtained via $T$-duality; 
here the 1-point functions are 
$$
 \langle \varphi_{g,\bar g}(z,\bar z) \rangle_{N\, \tilde x_0} 
  \ = \ \delta_{g,-\bar g}\; \sqrt{r}\;\frac { e^{2ir w \tilde x_0}} 
  {(z-\bar z)^{r^2w^2} } \ \  ;
$$ 
now we have $X(z) = \tilde x_0 + X(\bar z)$ for $z= \bar z$, 
where $\tilde x_0 \in \R\; {\rm mod}\; \frac{\pi}{r}$ parametrizes 
representations of the fundamental group $\pi_1(S^1)$ (`Wilson lines'). 
The $r$-dependent normalization arises from the non-trivial one-point 
function $\langle {\bf 1} \rangle_\a$. 

Passing to boundary states and applying a modular transformation as in 
(\ref{partfct}), we obtain the following formulas for the partition 
functions of the theories with boundary conditions $D\,x_0$ 
respectively $N\,\tilde x_0$ along the real line, 
\be
Z_{D\,x_0}(q) \ = \ \frac 1 {\eta(q) }\; \sum_{k\in\Z}  q^{2r^2k^2}  
  \ \ , \quad\quad
Z_{N\,\tilde x_0}(q)\  =\   \frac 1 {\eta(q) }\; \sum_{w\in\Z} 
 q^{\frac{w^2}{2r^2}}\ \ ; 
\label{freepartfcts}
\ee
they depend  on the compactification radius (i.e., on the bulk modulus), 
but not on $x_0$ or $\tilde x_0$. 

\subsubsection{Chiral deformations.}
Marginal deformations with the chiral current $\sJ(x)$ can be treated 
in close analogy to the uncompactified case: First observe that the 
gluing conditions are invariant under $\gamma_J$ and that only 
the coefficients $A_{ij}^{\alpha}$ can be effected by the perturbation. 
As before, the matrix $X^i_{J}$ acts on $\varphi_{g,\bar g}(z, \bar z)$ 
through multiplication by $g$, therefore eq.\ (\ref{form2}) leads to 
\ba
\langle \varphi_{g,\bar g}(z,\bar z) \rangle_{D\, x_0;\ \lambda J} 
\ & = &\ \delta_{g,\bar g}\;  \frac 1 {\sqrt{2r}} \;
 \frac {e^{i k (x_0+\frac{\lambda}2)/r}}
 {(z-\bar z)^{ \frac{k^2}{4r^2}} } \ \ , \nn\\[2mm] 
\langle \varphi_{g,\bar g}(z,\bar z) \rangle_{N\, \tilde x_0;\ \lambda J} 
\ & = &\ \delta_{g,-\bar g}\; \sqrt{r}\;
 \frac { e^{2ir w (\tilde x_0+\frac{\lambda}2)}}
  {(z-\bar z)^{r^2w^2} } \ \  :\nn
\ea
The marginal perturbations with the current $\sJ(x)$ induce translations 
in the Dirichlet resp.\ Neumann parameters $x_0$ and $\tilde x_0$ -- 
periodic in $\lambda$ with period $4\pi r$ resp.\ $2\pi/r$. 

For the ``rational radii'' $r= \sqrt{M/N}$ with positive coprime integers 
$M,N$, additional chiral (local) fields 
$$ W^\pm_{g_{{\rm loc}}}(z) \ = \ :\,e^{\pm i 2 g_{\rm loc} X_L(z)}\,:
  \ \ \ \mbox{ and }\ \ \  
   \bW{}^\pm_{g_{{\rm loc}}}(\bz) \ = \ :\,e^{\pm i 2 g_{\rm loc} X_R(\bz)}\,:
$$ 
(along with products) appear in the bulk theory; the charge $g_{{\rm loc}}$ is 
$2\sqrt{MN}$ if $N$ is odd and $\sqrt{MN}$ if $N$ is even.  These extended 
chiral algebras in the bulk theories are a well-known feature of the 
rational Gaussian models, see \cite{DVVV} and references therein. 
\newline
We may ask whether this additional symmetry is preserved by the boundary 
conditions and how the gluing conditions, if they exist, behave under 
marginal perturbations with the chiral current $\sJ(x)$. It is easily seen 
from the bosonization formula for $W_{g_{{\rm loc}}}^{\pm}(z)$ and 
$\bW{}^{\pm}_{g_{{\rm loc}}}(\bz)$ that all Dirichlet boundary theories 
respect the enhanced symmetry with gluing conditions  
\be
W^\pm_{g_{{\rm loc}}}(z) \ = \ \Omega_D\bigl[
  \overline{W}{}^\pm_{g_{{\rm loc}}}\bigr]
  (\bar z) \ := \ e^{\pm 2 i g_{{\rm loc}} x_0}\, 
  \overline{W}{}^\mp_{g_{{\rm loc}}}
  (\bar z)
\label{locfiegluconddir}
\ee
for $z = \bar z$. If the free boson satisfies Neumann boundary 
conditions,  $X_L(x) = X_R(x) + \tilde x_0$ leads to 
\be
W^\pm_{g_{{\rm loc}}}(z) \ = \ \Omega_N\bigl[\overline{W}{}^\pm_{g_{{\rm 
  loc}}}\bigr](\bar z) \ := \ e^{\pm 2 i g_{{\rm loc}} \tilde x_0}\, 
 \overline{W}{}^\pm_{g_{{\rm loc}}}(\bar z)
\label{locfieglucondneu}
\ee
along the boundary. Consequently, under marginal boundary perturbations 
with $\sJ(x)$, these enhanced gluing conditions are no longer invariant, 
instead they behave according to eq.\ (\ref{form1}) simply because 
$W^\pm_{g_{{\rm loc}}}$ are charged wrt.\ $\sJ$. 
\medskip

Something special occurs at the ``self-dual point'' $r=1/\sqrt{2}$: 
Here, the local chiral fields $J^{\pm}(z) := W^\pm_{\sqrt{2}}(z)$ 
and $\bJ{}^\pm (\bz) := \bW{}^\pm_{\sqrt{2}}(\bz)$ have conformal 
dimension $h_\pm = 1$.  This means that there are new marginal 
operators {\em within} the enlarged 
chiral algebra -- which is simply the  non-abelian current algebra 
${\rm SU}(2)_1$. We have seen that $J^{\pm}(z), \bJ{}^\pm(\bz)$ 
automatically obey the gluing conditions $\Omega$ from equation  
(\ref{locfiegluconddir}) or (\ref{locfieglucondneu}) for all 
Dirichlet or Neumann boundary theories. Therefore, SU(2)$_1$ is 
preserved at the boundary. 
\smallskip

The general results of Subsections 3.3 and 3.4 show that an arbitrary real 
linear combination of $\sJ^1 = \frac{1}{\sqrt{2}} (\sJ^+ + \sJ^-), 
\sJ^2 = \frac{1}{\sqrt{2}i} (\sJ^+ - \sJ^-)$ and $ \sJ^3 = \sJ $ 
can be used to deform the free bosonic boundary theories at 
$r_{\rm s.d.}$. The new models satisfy all sewing constraints and 
can be described by the boundary states 
\be
|\Gamma \rangle_{\rm s.d.} \ = \ \Gamma \, |N(0)\rangle_{\rm s.d.} 
\quad\quad
{\rm with} \ \ \Gamma \,= \, e^{i \sum \lambda_a J^a_0}\  
\label{su2fam}\ee
This family contains 
the boundary states $|N(\tilde x_0)\rangle$, but also other cases where 
$\sJ^3(z)$ does not obey simple Neumann gluing conditions. 

Naively, one might expect to obtain a second component of the moduli 
space of boundary theories by SU(2)-deformations of the Dirichlet 
boundary state $|D(0)\rangle_{\rm s.d.}$. However, the Dirichlet 
boundary states are already included 
in the set (\ref{su2fam}): By means of the SU(2)-deformations 
at the self-dual radius, we can rotate Neumann conditions for $J = 
J ^3$ into Dirichlet conditions; a perturbation with $\lambda\,\sJ^1$ 
changes the gluing condition $J^3(z) = \pm \bJ{}^3(\bar z)$ to 
$$
J^3(z) \ = \ \pm  (\cos \sqrt{2}\lambda)\, \bJ{}^3(\bar z)  
\pm  (\sin \sqrt{2}\lambda) \,\bJ{}^2(\bar z)  \ , 
$$
cf.\ the general formula (\ref{form1}) and also (\ref{newglcond}) for 
the non-chiral deformation $\psi^1$. When approaching $\lambda 
= \pi/\sqrt{2}$, Neumann conditions for $J^3$ turn into Dirichlet 
conditions -- by a continuous deformation. More precisely, we can write 
\be
|D(0)\rangle_{\rm s.d.} \ = \ e^{i \frac{\pi}{\sqrt2} J^1_0}\  
|N(0)\rangle_{\rm s.d.} \ , 
\label{DirNeuRsd}\ee
showing that there is one connected SU(2)-family of boundary 
conditions for $r_{\rm s.d.}$. 
\smallskip

Let us compare the structure of 
boundary theories at the self-dual radius to the known boundary 
states of the  $SU(2)_1$ WZW-model \cite{Ish1,Car3}. The possibility 
of non-standard gluing conditions for the currents was not realized in 
these works, but with the help of the general formalism explained in 
Section 2 it is straightforward to extend Cardy's classification of 
boundary 
states to arbitrary gluing maps $\Omega$ in the SU(2) current algebra.  
Per fixed gluing condition $\Omega$, one finds two boundary states 
$$  |i\rangle_{\Omega} \ = \ 2^{1/4}\, |0\rangle\!\rangle_{\Omega} 
      + (-1)^i\, 2^{1/4}\, |1\rangle\!\rangle_{\Omega}
$$
where $i=0,1$ labels the two irreducible heighest weight 
representations of SU(2)$_1$. To re-discover those in the family 
(\ref{su2fam}), observe that formula (\ref{locfieglucondneu}) 
with $g_{\rm loc}= \sqrt{2}$ is 
invariant under the shift $x_0 \longmapsto x_0 + \pi/\sqrt{2}$, 
while the marginal perturbation $\exp(-i \sqrt{2}\pi J^3_0)$ 
implementing this shift acts non-trivially on the full boundary 
state $|N(\tilde x_0)\rangle$, producing precisely the sign in 
front of the spin $1/2$ Ishibashi state $|1\rangle\!\rangle_{N}$. 
Thus, $|N(0)\rangle$ and $|N(\pi \sqrt{2})\rangle$ -- sitting 
at opposite points of the $\tilde x_0$-circle -- coincide with 
Cardy's rational SU(2)$_1$ boundary states. Analogous results 
hold for other gluing conditions, which  
are parameterized by SO(3) since central elements of SU(2) yield 
trivial $\gamma_J$ in eq.\ (\ref{form1}). But there are two 
different boundary theories sitting over each point of this 
SO(3) which resolve the full SU(2) moduli space we found before.  
Cardy's boundary states for the SU(2)$_1$ WZW-model are simply 
assigned to elements in the centre of SU(2).  
\medskip

All the time, we have implicitly assumed that the boundary 
conditions in (\ref{su2fam}) are pairwise inequivalent -- which 
is not clear a priori. In the self-dual bulk theory, e.g., all 
operators $J'(z)\overline{J}{}'(\bar z)$ with $J'=\sum \lambda_a J^a$ 
are marginal 
and we can move away from $r=1/\sqrt{2}$ along an $S^3$ of different 
directions. But all these deformations result in equivalent bulk CFTs 
because of SU(2)$\,\times\,$SU(2) symmetry, leaving only the ordinary 
change-of-radius deformation. 

For boundary CFTs, we arrive at a similar scenario if we declare 
boundary conditions $(\Omega_i, \alpha_i)$, $i = 1,2$, equivalent as 
soon as there is an automorphism (a ``gauge transformation'') of the 
bulk CFT  which intertwines the gluing conditions 
$\Omega_1$ and $\Omega_2$ and maps the set of 1-point-functions 
$A^{\a_1}$ to $A^{\a_2}$. This criterion, however,  would even identify 
all possible Dirichlet conditions $(D, x_0)$ for a free boson, simply 
because of translational invariance. 

We can formulate a sharper criterion by composing new 
systems from two different boundary conditions, e.g.\ by putting 
the CFT on the strip with boundary conditions $\alpha$ on one end 
and $\beta$ on the other. Then, additional data like the  partition 
function $Z_{\a\b}(q)$ discussed in Section 3.4 are available, and 
we can certainly conclude 
that $\alpha \not\simeq \beta$ if $Z_{\a\a}(q) \neq Z_{\a\b}(q)$. 

In this way, not only can we resolve all the free boson boundary conditions 
at generic radii, but also the family  (\ref{su2fam}) at the self-dual 
point. Since every SU(2)$_1$ boundary state
can obtained from one out of Cardy's list by the action of an  SU(2) 
element $g = \exp(i J'_0)$, we have to 
compute $Z_{\a\b}(q)$ for some $|\alpha\rangle$ (which obeys, say, standard 
gluing conditions) and arbitrary $|\b\rangle := g|\alpha\rangle$. 
First note that $Z_{h\a, h\b}(q) = Z_{\a,\b}(q)$ and 
$Z_{\a, hgh^{-1}\b}(q) = Z_{\a,g\b}(q)$ for all 
$g,h\in\, $SU(2) -- this follows from $h\,|\alpha\rangle = 
\bar h^{-1} |\alpha\rangle$ and $\Theta h = h \Theta$. Therefore, 
$Z_{\a,g\a}(q)$ depends only on the conjugacy class of $g$, and we 
can in particular choose an $h$ such that $h g h^{-1} = t \equiv
\exp{(i\lambda J^3_0)}$ is in a given torus of SU(2). 

Partition functions  $Z_{\a\b}(q)$ where one of the boundary states 
has been twisted by a current in a maximal abelian subgroup can be 
computed with standard modular transformation rules. If $\alpha$ is 
one of Cardy's boundary conditions, we find the expression 
$$
Z_{\a,t\a}(q) = \langle \Theta \a|\, \tilde q^{L_0 - \frac {c}{24}} 
e^{i\lambda J^3_0}\, |\alpha\rangle = 
\sum_i N_{\a, \a^+}^i \, {\rm tr}_{{\cal H}_i} 
q^{L_0 + \frac{\lambda}{2\pi} J^3_0 +  \frac{\lambda^2}{8\pi^2} - 
\frac {c}{24}} \ , 
$$
which involves twisted SU(2)-characters that depend on $\lambda$. \newline
Finally, if $g \neq g'$ are conjugate to the same torus element, one 
can show that there is a boundary condition $\a'$ such that 
$Z_{\a', g\a}(q) \neq Z_{\a', g'\a}(q)$ -- yielding a complete 
resolution of the SU(2)-family  (\ref{su2fam}) as desired.

\subsubsection{Non-chiral deformations.} At special values of the 
compactification radius, there are extra non-chiral marginal deformations 
similar to the ones present for the uncompactified 
free boson with Neumann boundary condition. 
The partition functions (\ref{freepartfcts}) show that these radii are 
$r = N/ \sqrt{2}$ for integer $N$ in the Neumann and $r = 1/(\sqrt{2} N)$
in the Dirichlet case -- in accordance with an interpretation of 
the perturbation as a periodic boundary potential with period $1/\sqrt{2}$.
\newline  
The two self-local and mutually local primaries $\psi^{a}(x)$, 
$a= 1,2$, appearing there lead to similar effects as the non-chiral 
marginal operators in the uncompactified theory with Neumann boundary 
conditions. For $r = N/ \sqrt{2}$, the decomposition of the bulk Hilbert 
space into Virasoro modules results in a formula analogous to 
(\ref{flatbulkVir}); we write it in the form 
 $$
\cH^{(P)}  \ =  \ \bigoplus_{k,w \in \Z} 
\cV^{\rm U(1)}_{{k\over \sqrt{2} N} + {wN\over\sqrt{2}}}
\otimes  \cV^{\rm U(1)}_{{k\over \sqrt{2} N} - {wN\over\sqrt{2}}}
 \ \oplus \ldots \ =\ \bigoplus_{j\in{\oh}\Z_+} \;
\bigoplus_{k',w \in \Z} 
\cV^{{\rm Vir} \times {\rm Vir}}_{{j^2, {k'+Nw\over2},{k'-Nw\over2} }} \ 
\oplus \ \ldots
$$
where again the dots indicate terms that do not couple to the 
boundary, either because $h\neq \bar h$ or because the charge 
condition $g + \bar g \in \sqrt{2} \Z$ is not met. We have 
indicated the SU(2) quantum numbers explicitly, adopting the 
convention that a module ${\cal V}_{j, m,n}$ is empty unless 
$m$ and $n$ are in the range $-j, \ldots, j$. Precisely the 
same Virasoro primaries contribute if we consider the 
perturbation of a boundary CFT with Dirichlet conditions at 
radius $r = \frac{1}{\sqrt{2}N}$. 

We can apply the methods used in the uncompactified case 
to determine the deformed boundary states, and we find 
\be
|N(\tilde x_0);\,\lambda \psi^a\rangle \ =\ 
\sum_{j\in{\oh}\Z_+} \sum_{w, k' \in \Z} 
D^{\ j}_{{k'+Nw\over2}, {-k'+Nw\over2}} (\Gamma^a_{\tilde x_0,\lambda}) \; 
|j, {\scriptstyle {k'+Nw\over2}}, { \scriptstyle {k'-Nw\over2}} 
\rangle\!\rangle\ .
\label{comppertbdst}\ee
Now, the SU(2)-element $\Gamma^a_{\tilde x_0,\lambda} = 
e^{i \lambda \psi^a_0} e^{2i \tilde x_0 J_0}$  contains the 
perturbation parameter $\lambda$ along with $\tilde x_0$ specifying 
the original Neumann condition. The latter is recovered for $\lambda=0$, 
where only the terms with $k'=0$ contribute (the $N$-dependence encodes 
the information on the radius). 
\smallskip

Again, the modular transformation to obtain the spectrum from the boundary 
states is not manageable except for the Dirichlet-like points 
$\lambda={2k+1\over \sqrt{2}}\pi$ (they are  ``Neumann-like'' points if 
we start from Dirichlet conditions at the dual radius). There, the 
prefactors of the Virasoro Ishibashi states are given by  the phases 
$$
D^{\;j}_{{k'+Nw\over2},{-k' + Nw\over2}}
(\Gamma^1_{\tilde x_0,\lambda_{{\rm Dir}}}) 
\ = \ \delta_{w,0} (-1)^j e^{-\sqrt{2} i \tilde x_0 k'}\ \ , $$
which lead to the same perturbed partition function (\ref{Zdirlike}) 
as in the uncompactified case. In particular, the parameter $\tilde x_0$ 
does not appear in $Z_{\alpha_D}(q)$, and in the boundary state itself 
it shows up with a different periodicity: The information about the 
original radius $r={N\over\sqrt{2}}$ has been lost during the 
perturbation. 
\medskip

The alternative method of \cite{PoTh} applies again, and it leads
to a formula for the partition function similar to eq.\ (\ref{ZPolTh}), 
only that the $\zeta$-integral is to be replaced by a sum since the 
spectrum is discrete from the start. 
For later purposes, let us give the explicit formula for the case 
$r = \sqrt{2}$: 
With $|\alpha\rangle_\lambda := |N(\tilde x_0);\,\lambda \psi^a\rangle$, 
restriction of the $\zeta$-integral in (\ref{ZPolTh}) to the sum over 
$0, {\scriptstyle {\oh}}, 1, {\scriptstyle {3\over2}}$ yields 
\be
Z_{\alpha_\lambda}(q) 
= \eta(q)^{-1}\;\sum_{m\in\Z} \, \Bigl(\, q^{m^2} +   
             q^{(m + \oh + {\lambda \over \sqrt{2}\pi} )^2} \Bigr) \ .
\label{compZPolTh}\ee
Generally, the charges $g\neq n \sqrt{2}$ 
follow the flow prescribed by the function (\ref{arcsinfunc}), 
the corresponding fields being those which are non-local 
wrt.\ to the perturbing field. Finally, only charges  $g = n 
\sqrt{2}$ are left at $\lambda = {\pi\over \sqrt{2}}$. It is 
once more easy to show that the cluster property is broken 
at the Dirichlet-like point, but we have no direct handle on 
clustering for intermediate $\lambda$. Employing the higher symmetry 
algebras present at rational radii does not seem to yield additional
insight into the clustering properties, either. Surprisingly, however, 
the study of orbifold models will provide further information. 

\medskip

\subsection{The {\fatma c}$\;$=$\;$1 orbifold theories.} 
The moduli space of $c=1$ theories on the plane has another branch which 
parameterizes orbifolds of the circle theories. This family is constructed 
by ``dividing out'' the left-right symmetric $\Z_2$-action $X \longmapsto 
-X$ on the compactified free boson theories -- see e.g.\ \cite{DVVV,Gin1} and 
references therein. The chiral fields are the invariant elements of 
the U(1)$\,\times\,$U(1) current algebra, the bulk Hilbert space consists of 
an untwisted sector containing all $\Z_2$-invariant states of the free 
boson Hilbert space and of two twisted sectors $\cH^{\rm tw}_{0}$ and 
$\cH^{\rm tw}_{\pi r}$ built up over twist fields of left and right 
conformal dimension $h^{\rm tw}_{0,\pi r} = 1/16 $. The subscripts 
refer to the endpoints of the interval $[0,\pi r]$ which can be 
regarded as the target space of the orbifold model at radius $r$. 
For $r=r_{\rm s.d.}$, there are three further orbifold models that 
arise from dividing out finite subgroups of SO(3), see \cite{Gin2,DVVV}, 
but we will not discuss these cases here.  

We give the description of the associated boundary orbifold models in 
terms of boundary states, which can e.g.\ be found in \cite{OsAff}.
Consider the untwisted sector first. The free boson Ishibashi states 
are given as $\Z_2$-invariant exponentials of $\sum a_{-n}\bar a_{-n}$ 
acting on U(1) ground states; therefore one merely has to symmetrize 
in the latter to obtain ``untwisted'' orbifold  boundary states from 
ordinary free boson Dirichlet or Neumann boundary states, 
\ba
|D(x_0)\rangle^{\rm orb} & := & \frac1{\sqrt2}\,\Bigl(\; 
  |D(x_0)\rangle^{\rm circ} + |D(-x_0)\rangle^{\rm circ} \;\Bigr)\ ,
\label{orbDir}\\[2mm]
|N(\tilde x_0)\rangle^{\rm orb} & := & \frac1{\sqrt2}\,\Bigl(\;
  |N(\tilde x_0)\rangle^{\rm circ} + |N(-\tilde x_0)\rangle^{\rm circ}
  \;\Bigr)\ ; \label{orbNeu}
\ea
The parameters range over the intervals $0< x_0 < \pi r$ and 
$0< \tilde x_0 < \frac {\pi} {2 r}$. In terms of 1-point functions, 
(\ref{orbDir}) e.g.\ means that 
$$
\langle \, \cos \bigl({\textstyle \frac{k}{r}} X(z,\bar z)\bigr)\, 
  \rangle^{\rm orb}_{D\, x_0} \ = \ \frac 1 {\sqrt{2r}} \;
  \frac{\cos \frac{kx_0}{r}}{(z-\bz)^{k^2/4r^2}}
$$ 
and that no twist fields couple to the 
identity on the boundary. A similar formula holds for Neumann boundary 
conditions of the orbifold theory. 
\smallskip
       
To each fixed point of the $\Z_2$-action on $S^1$, one assigns two 
{\em twisted} Dirichlet and two {\em twisted} Neumann boundary states 
made up from the corresponding circle boundary states and the 
(appropriately symmetrized) Dirichlet or Neumann Ishibashi states of 
$\cH^{\rm tw}_{0,\pi r}$, see \cite{OsAff} for more details. 
With $\xi = 0, \pi r$ and $\tilde \xi = 0, 
\frac{\pi}{2r}$, we write 
\ba
|D(\xi), \pm \rangle^{\rm orb} & := & 2^{-\frac1 2}\,
  |D(\xi)\rangle^{\rm circ} \pm 2^{-\frac1 4}\, |D(\xi)\rangle^{\rm tw} \ ,
\label{orbtwDir} \\[2mm]
|N(\tilde\xi), \pm \rangle^{\rm orb} & := & 2^{-\frac1 2}\,
  |N(\tilde\xi)\rangle^{\rm circ} \pm 2^{-\frac1 4}\,  
  |N(\tilde\xi)\rangle^{\rm tw} \ .
\label{orbtwNeu}
\ea
The prefactors ensure proper normalization of all partition functions 
$Z_{\a\b}(q)$ for $\a,\b$ taken from the two sets (\ref{orbDir}-\ref{orbtwNeu}). 
For our purposes, 
the cases with $\a=\b$ are most important since they provide 
the number of marginal boundary operators induced by the boundary 
condition $\alpha$. In the case of Dirichlet gluing conditions, one 
obtains 
\ba
Z_{\a}(q) & = &  \sum_{k\in\Z}\; \frac {q^{2r^2k^2}}{\eta(q)}  + 
 \sum_{k\in\Z}\; \frac {q^{2(rk+\frac{x_0}{\pi})^2}}{\eta(q)}\quad 
\ \ \mbox{\rm for}\ \ |\a\rangle \ = \ |D(x_0)\rangle^{\rm orb} \ ,
\label{Zuntw}\\[2mm]
Z_{\b}(q) & = &  \sum_{k=1}^{\infty}\; \frac {q^{2r^2k^2}}{\eta(q)}
+ \sum_{n=0}^{\infty} \ \chi^{\rm Vir}_{4n^2}(q)\quad \ \ \ 
\mbox{\rm for}\ \ \ |\b\rangle \ =\ | D(\xi), \pm \rangle^{\rm orb}\ ;
\label{Ztw}
\ea
the Neumann partition functions follow when $r$ is replaced with $1/2r$. 
The Virasoro characters $\chi^{\rm Vir}_h(q)$ were introduced in Subsection 
4.1.2.  They coincide with $\eta^{-1} q^h$ if $h\neq 
m^2$ for any $m\in \frac12 \Z$, and are given by the difference
(\ref{excepVirchars}) of U(1) characters otherwise.  
\smallskip

Since the U(1) current algebra is reduced by the orbifold procedure, 
the occurrence of Virasoro characters for twisted boundary states 
is not surprising. Indeed, (\ref{Ztw}) is precisely the $\Z_2$-projection 
of the circle Dirichlet partition function, the second sum being 
the vacuum character of the $\Z_2$-invariant subalgebra of U(1). 

On the other hand, the partition functions for untwisted Dirichlet boundary 
conditions (\ref{orbDir}) are sums of U(1) characters; the state 
space of the corresponding boundary theories is {\em not} $\Z_2$-invariant,  
and (\ref{Zuntw}) should be interpreted as the total excitation spectrum of 
a superposition of two branes in the circle theory. (Nevertheless, 
the boundary states above obey the cluster property with respect to 
the reduced set of bulk fields present in the orbifold theory.) The first 
sum in (\ref{Zuntw}) describes strings starting and ending on the same 
brane, whereas the $x_0$-dependent 
characters are associated with excitations of open strings stretching 
between the Dirichlet brane at $x_0$ to the one at $-x_0$, up to 
identification of strings running in opposite directions. The corresponding 
boundary fields are induced by the bulk-boundary OPE of the twist fields
$\sigma_{0,\pi r}(z,\bz)$ in the bulk \cite{OsAff}. 
\smallskip

The marginal boundary operator content of the orbifold models, too, 
depends on $r$ and $x_0$. Let us look at untwisted Dirichlet boundary 
conditions first (always, the Neumann cases follow upon $T$-dualizing 
the radius): For arbitrary radius $r$, one marginal operator $J(x)$ 
occurs in the parameter-independent part of the partition function 
for arbitrary radius $r$, in the vacuum U(1) character. This field 
is the boundary value of the original bulk current of the circle theory 
which was removed by the orbifolding procedure, and it appears through 
the bulk-boundary OPE of the bulk fields $\cos(\frac{k}{r}X)$ with 
$k\neq 0$, 
\be  \cos \bigl({\textstyle \frac{k}{r}} X(z,\bar z)\bigr) \ = \ 
    \frac{\cos \frac{kx_0}{r}}{(z-\bz)^{\frac{k^2}{4r^2}}}\ {\bf 1}  \ - \ 
    \frac{i\, \frac{k}{2r}\, \sin \frac{kx_0}{r}}{(z-\bz)^{\frac{k^2}{4r^2} 
    - 1 }}\ J(x) \ + \ \dots   \label{bbOPEcos} \ee 
$J(x)$ is local with respect to all other boundary fields from the 
$x_0$-independent part of the spectrum, but non-local wrt.\  
those fields which have an $x_0$-dependent conformal dimension, 
since the latter arise through the bulk-boundary OPE of twist 
fields. Consequently, the second part of the boundary spectrum 
(\ref{Zuntw}) is not protected against changes under a perturbation
with $J$. This is perfectly consistent with our findings below that
$J$ simply moves the position $x_0$ of the brane.  

For the special radii $r=\frac 1 {\sqrt{2}N}$, two additional 
states $\psi^{a}(x),\ a=1,2$, of dimension 1 show up in the 
$x_0$-independent part of the partition function (\ref{Zuntw}). 
They are self-local and give rise to the familiar periodic boundary 
potentials. 

The parameter-dependent part of $Z_{\a}(q)$ can contain further 
marginal operators if the distance of the two branes -- the 
length of the stretched open string -- 
is appropriately 
adjusted: If $r= 1 /(\sqrt{2} N)$ with $N\in \Z$, this fine-tuning 
cannot be achieved, but for all other radii there is one marginal 
field $\psi'(x)$ whenever  $x_0 = 1/ \sqrt2 - k_0 r\;$\ or\  $\;x_0 = 
- 1 / \sqrt2 + (k_0+1)\, r$, where $k_0$ is the positive integer 
satisfying $\sqrt{2} r k_0 < 1 < \sqrt{2} r (k_0+1)$. Since 
these massless excitations originate from the 
bulk-boundary OPE of a twist field in the bulk, they will 
have non-trivial monodromy wrt.\ the twist field and wrt.\ themselves, 
hence they are non-local and do not give rise to analytic deformations.     

The picture is simpler for twisted boundary conditions: 
There is no field of dimension $h=1$ in the  $\Z_2$-invariant 
subalgebra of the U(1) current algebra, and the first sum in 
(\ref{Ztw}) contributes one marginal operator iff $r=1/(\sqrt{2}N)$; 
this is just the boundary field $\sqrt{2}\,\cos(2\sqrt{2} X(x))$, 
leading to similar effects as $\psi^a(x)$.  
\medskip

When constructing the deformed boundary theories, one encounters 
the same general phenomenon as for the unorbifolded models: Some 
of the boundary conditions listed above are connected by boundary 
perturbations and, at special values of the bulk parameters, new 
boundary states are generated that would have been hard to discover 
directly without using marginal deformations. 
\newline
Let us first  focus on the perturbation of the untwisted boundary 
states generated by the self-local marginal field $J(x)$. This 
deformation does not change the 
Dirichlet or Neumann gluing conditions of the orbifold theory. 
Furthermore, since  $J(x)$ was defined through the bulk-boundary 
OPE (\ref{bbOPEcos}) of a bulk field from the untwisted sector, 
we conclude that the 1-point functions of bulk twist 
fields continue to vanish in the $J$-deformed theory. 
To calculate the effect on the 1-point functions of untwisted fields, 
we use (\ref{orbDir}) to pass to the underlying circle theory, 
where the deformation by a current is easy to handle. However, 
observe that the coefficient of $J(x)$ in the bulk-boundary 
OPE (\ref{bbOPEcos}) is antisymmetric upon replacing $x_0$ by 
$-x_0$, so the definition of the current $J$ picks up an extra 
minus sign when acting on the second term in the boundary
state (\ref{orbDir}). The result is that, as long as $0 <  
x_0+{\textstyle\frac{\lambda}2} < \pi r$,     
\be
|D(x_0)\rangle^{\rm orb}_{\lambda J}  \ = {\textstyle \frac 1{\sqrt{2}}}\, 
  e^{i \lambda J_0} 
  \ |D(x_0)\rangle^{\rm circ} + {\textstyle \frac 1{\sqrt{2}}} \, 
  e^{-i\lambda J_0} \ |D(-x_0)\rangle^{\rm circ} \ = \   
             |D(x_0+{\textstyle\frac{\lambda}2})\rangle^{\rm orb}\ .
\label{orbDirdef} 
\ee
The marginal operator $J(x)$ moves the untwisted orbifold brane along 
the interval $]0,\pi r[$. Continuation into the end-points $\xi$ leads to 
the boundary states $|D(\xi),+\rangle^{\rm orb}+|D(\xi),-\rangle^{\rm orb}$, 
which are inconsistent in the sense that 
they violate the sewing relation (\ref{class}) for the twist fields. 
In the interior of the interval, however, the deformed theory has the cluster 
property in spite of being generated by a non-chiral deformation, and 
the spectrum behaves as expected.

The perturbations with $\psi^a(x)$ from untwisted or with the marginal 
operator from twisted boundary conditions have to be treated in analogy to 
the unorbifolded case, and the technical details were provided in 
Subsections 4.1.2 and 4.2.2.  Let us, however, have a closer look at 
the radius $r = 1/\sqrt{2}$, which is again exceptional. Among the bulk 
fields, there is one chiral current, $J^1_{\rm orb}(z) :=\sqrt{2}\, 
\cos 2\sqrt{2} X(z)$, and it is easy to see that some of the boundary 
conditions (\ref{orbDir}-\ref{orbtwNeu}) preserve this extended symmetry: 
\newline
$J^1_{\rm orb}(z)$ satisfies Dirichlet gluing conditions 
for $|D(x_0)\rangle^{\rm orb}$ or $|N(\tilde x_0)\rangle^{\rm orb}$
if $x_0=\tilde x_0 = \frac{\pi}{2\sqrt{2}}$. The eight twisted boundary 
states (\ref{orbtwDir},\ref{orbtwNeu}) induce Neumann boundary conditions 
on $J^1_{\rm orb}(z)$. 
\newline
In those cases,  $J^1_{\rm orb}$ is a chiral 
local field of the full boundary CFT, and it follows from the general theory 
developed in Subsections 3.3, 3.4 that the boundary value $J^1_{\rm orb}(x)$ 
generates deformations which neither change the spectrum nor violate clustering 
conditions. We obtain two continuous U(1) families of deformed boundary states, 
containing the two untwisted resp.\ the eight twisted boundary states from 
above which exist at generic radii. The first family is further enlarged by 
$J_{\rm orb}$, see also below.

In the bulk, the $r=1/\sqrt2$ orbifold model is equivalent to 
the $r=\sqrt2$ circle theory, see  e.g.\ \cite{Gin1}: The identification of the 
two models starts from the $r = 1/\sqrt{2}$ circle theory, where the two 
different orbifoldings $X \sim -X$ (i.e.\ $J^3 \sim -J^3$) and 
 $X \sim X + \frac{2\pi r}{2}$ (i.e.\ $J^1 \sim -J^1$) are equivalent
by SU(2)-symmetry; the second procedure leads to a circle model at $r=1/\sqrt{8}$, 
which in turn is $T$-dual to the  $r=\sqrt{2}$ theory. 

It is quite instructive to investigate how this equivalence relates 
boundary conditions for the bulk theories, so we give an outline. 
The chain of isomorphisms sketched above implies that Dirichlet resp.\ 
Neumann gluing conditions for $J^3_{\rm circ}$ in the $r_{\rm circ}= \sqrt{2}$ 
model correspond to Neumann resp.\ 
Dirichlet conditions for $J^1_{\rm orb}$ in the $r_{\rm orb} = 1/\sqrt{2}$ 
theory. We have already singled out the latter orbifold boundary states, 
and the following partition functions indeed coincide: 
\ba
Z_{\a}^{\sqrt{2}} (q) &=& Z_{\b}^{1/\sqrt{2}} (q) 
\quad\ \ {\rm for}\ \ |\a\rangle = |N(\tilde x_0)\rangle^{\rm circ}\ , \ \ 
|\b\rangle = |G({\textstyle \frac{\pi}{2\sqrt{2}}})\rangle^{\rm orb}\ ,
\phantom{xxx}\label{Zciruntw} \\
Z_{\a}^{\sqrt{2}} (q) &=& Z_{\b}^{1/\sqrt{2}} (q) 
\quad\ \ {\rm for}\ \ |\a\rangle = |D(x_0)\rangle^{\rm circ}\ , \ \ 
|\b\rangle = |G(\xi),\pm\rangle^{\rm orb}\ ; \phantom{xxx}
\label{Zcirtw}\ea
the gluing conditions ``$G$'' in the orbifold theory can be both $N$ or $D$, 
and the circle parameters take values $x_0 \in [\,0,2\sqrt{2}\pi]$ and 
$\tilde x_0 \in [\,0, \pi / \sqrt{2}]$ as usual. 
It is possible to pin down the one-to-one equivalence of boundary states by 
comparing the 1-point functions of corresponding bulk fields from 
circle and orbifold model; e.g., the twist fields $\sigma_0$ and 
$\sigma_{\pi r}$ are to be identified with $\sin(\frac1{\sqrt{2}} X)$ 
and  $\cos(\frac 1{\sqrt{2}} X)$ in the $r_{\rm circ}= \sqrt{2}$ theory 
on dimensional grounds. We restrict ourselves to some general observations: 

As $\tilde x_0$ in (\ref{Zciruntw}) is varied by the deformation 
$\lambda J^3_{\rm circ}$, 
the corresponding operator $\lambda J^1_{\rm orb}$ generates the  
U(1) family of orbifold boundary states mentioned above, with Dirichlet 
and Neumann gluing conditions for $J^3_{\rm orb}$ showing up at the opposite 
points $\lambda = 0$ and  $\lambda =  \pi / \sqrt{2}$ (compare the 
discussion of the self-dual circle model).

The twisted boundary 
states in (\ref{Zcirtw}), too, are members of a family generated by 
$J^1_{\rm orb}$. The identification of twist fields with vertex operators 
of the circle theory shows that under this deformation -- resp.\  under 
the $J^3_{\rm circ}$-perturbation -- the 1-point functions of $\sigma_0$ and 
$\sigma_{\pi r}$ can be turned on and off smoothly. We may say that 
$J^1_{\rm orb}$ induces a {\em tunneling} of the twisted D-brane states 
between the two $\Z_2$-fixed points. 

Let us try to match the ``missing'' orbifold boundary states, namely 
(\ref{orbDir},\ref{orbNeu}) with $x_0, \tilde x_0 \neq \frac{\pi}{2\sqrt{2}}$, 
to boundary conditions of the circle model. 
The isomorphism from the $r_{\rm orb} = 1/\sqrt{2}$ to the $r_{\rm circ} =
\sqrt{2}$ theory not only maps $J^1_{\rm orb}(z)$ to $J^3_{\rm circ}(z)$, 
but also allows us to identify the non-chiral boundary field 
$J^3_{\rm orb}(x) := J_{\rm orb}(x)$ with  $\psi^2_{\rm circ}(x)$ and 
$J^2_{\rm orb}(x) := \sqrt{2}\sin(2\sqrt 2 X(x))$ with $\psi^1_{\rm circ}(x)$. 
The orbifold boundary states in question are 
generated by $J_{\rm orb}(x)$ and do not preserve the $J^1_{\rm orb}$-symmetry. 
Likewise, the $\psi^2_{\rm circ}(x)$-deformed boundary states 
$|N(\tilde x_0);\, \lambda \psi^2\,\rangle^{\rm circ}$ of the 
circle model break the  $J^3_{\rm circ}$ gluing conditions. 
Furthermore, eqs.\ (\ref{Zuntw}) and (\ref{compZPolTh}) show that the
following  partition functions coincide, 
\be
Z_{\a}^{\sqrt{2}} (q) = Z_{\b}^{1/\sqrt{2}} (q)  \ \ \quad {\rm for}\ \ 
|\a\rangle = |N(\tilde x_0);\, \lambda \psi^a\,\rangle^{\rm circ} \ , \quad 
|\b\rangle = |G(x_0')\rangle^{\rm orb} 
\label{partfctmatch}\ee 
if $x_0' = \frac{\pi}{2\sqrt{2}} + \frac{\lambda}2\,$. 
All this tells us that the family of orbifold boundary states 
generated from $|N(\frac{\pi}{2\sqrt{2}})\rangle^{\rm orb}$ 
by $J^a_{\rm orb}$, $a=1,2,3$, 
corresponds to the family of circle boundary states generated from 
 $|N(0)\rangle^{\rm circ}$ by $J^3_{\rm circ}$ and $\psi^a_{\rm circ}$.

Because of the degeneracy in the partition functions, (\ref{partfctmatch})
does not quite allow us to match individual members of the families, and 
a direct comparison of 1-point-functions is virtually impossible because 
of the complicated matrix elements $D^{\,j}_{mn}(\Gamma)$ 
in eq.\ (\ref{comppertbdst}). 
Still, we can now draw general conclusions on the $\psi^a_{\rm circ}$-deformed 
boundary conditions of the $r_{\rm circ} =\sqrt{2}$ circle model 
that were inaccessible before: 

Perturbations by $\lambda \psi^a_{\rm circ}$ do preserve the cluster 
property for $|\lambda| <  \frac{\pi}{\sqrt{2}}$ since the corresponding 
orbifold boundary conditions do. It follows that the subfamilies 
of boundary conditions generated by $\psi^1$ or $\psi^2$ form 
open intervals. Altogether, $J^3_{\rm circ}$ and $\psi^a_{\rm circ}$ 
generate a patch of moduli space with the topology of the interior 
of a solid 2-torus (of a ``bagel''), which can be seen as follows: 
As long as we ignore clustering issues, these marginal operators lead to 
an SU(2) $\simeq S^3$ of boundary conditions when applied to 
$|N(0)\rangle^{\rm circ}$. We have to remove all points 
where clustering is violated -- which are characterized by Dirichlet gluing 
conditions for $J^3_{\rm circ}$.  The latter are broken by any infinitesimal 
perturbation with $\psi^a_{\rm circ}$, but $J^3_{\rm circ}$ itself maps 
the Dirichlet-like points into each other. Therefore, the 
remaining space of clustering boundary conditions is the bagel $S^3 \setminus 
S^1$. 

A direct isomorphism between orbifold and circle model can be exploited 
only for $r_{\rm circ} =\sqrt{2}$. Nevertheless, we expect the same  
topology to arise from the non-chiral perturbations at other radii 
$r_{\rm circ} =N / \sqrt{2} $, and a similar one in the uncompactified 
case (see below). As we have argued before, the 
breakdown of cluster properties at finite perturbation strength 
$\lambda = \pi / \sqrt{2} $ in $\lambda \psi^a(x)$ should be
due to a finite domain of convergence in the proof of clustering 
mentioned in Subsection 3.5.3.

\section{The {\fatma c} =1 brane moduli space, string geometry, and open problems}

Putting together the pieces found in the last section, we can give 
a global description of the moduli 
space  of $c=1$ conformal boundary conditions. This is possible 
because we could analyse marginal deformations 
to all orders in the perturbation parameter; first order results would 
have allowed for a local picture only. 
\newline
The (brane) moduli space $\cM_B$ can be viewed as a fibration 
over the (closed string) moduli space $\cM_S$ of bulk 
CFTs, $ \cM_B = \bigcup_{m\in \cM_S} \bigl(\cM_B\bigr)_m\,$. We focus 
on the connected part  $\cM_S = \cM_S^{\rm circ} \cup \cM_S^{\rm orb}$  
and ignore the three exceptional orbifold points. Both 
branches of $\cM_S$ are parameterized as half-lines $\R_{\geq 1/\sqrt{2}}$, 
since radii below the self-dual one lead to equivalent theories upon 
$T$-duality $r \leftrightarrow 1/2r$ and exchange of Dirichlet and Neumann 
boundary conditions.

The topological 
type of the fiber $\bigl(\cM_B\bigr)_m$ depends on $m$: 
For $m = r_{\rm circ} \in \cM_S^{\rm circ}$, we found 
\def\disju{\cup\hskip-8.5pt\cdot\hskip8.5pt}
\be
\bigl(\cM_B\bigr)_{r_{\rm circ}} \ = \ \left\{  \begin{array}{ll}
S^1_r \disju S^1_{1/{2r}}  &\ \  r_{\rm circ} \neq \frac N{\sqrt{2}} \\[2mm] 
S^1_r \disju  B_{1/2r} &\ \   r_{\rm circ} = \frac N{\sqrt{2}}\,,\ N \geq 2   \\[2mm]
S^3 &\ \  r_{\rm circ} =  \frac 1{\sqrt{2}}   \\[2mm]
\R \disju \widetilde{B} &\ \ r_{\rm circ} = \infty \ \; .
        \end{array} \right. 
 \label{modspacecirc}
\ee
Points $x_0$ in $S^1_r$ label positions of Dirichlet branes, while the 
Neumann parameter $\tilde x_0 \in S^1_{1/{2r}}$ distinguishes Wilson lines. 
\newline
The spaces 
$B_{1/2r} \simeq \mbox{{\it \r{D}}}\,{}^2_{\pi/\sqrt{2}} \times S^1_{1/2r}$ 
have the topology of the soft interior of a bagel before baking,  
cf.\ the end of Subsection 4.3. The boundary of the 2-disk 
$D^2_{\pi/\sqrt{2}}$ corresponds to Dirichlet-like mixtures of pure 
boundary conditions, which violate the cluster property.
\newline
The uncompactified case  emerges in the $N \ra \infty$ limit of the 
second line in (\ref{modspacecirc}): The component $\R$ indicates 
that the brane can be placed anywhere in the flat target. The second 
component $\widetilde{B} \simeq 
B_{1/2r}\,/(0 \times S^1_{1/2r}) $ has the topology of an open solid torus with  
the central circle shrunk to a point. This can be seen from the matrix 
elements $D^{\,j}_{m,-m} (\Gamma^a_{\tilde x_0, \lambda})$ which, for 
$\lambda = 0$, become independent of the parameter $\tilde x_0$ -- 
in agreement with the fact that $\R$ is simply connected. Switching 
on a periodic boundary potential, however, lifts the $\tilde x_0$-degeneracy. 

Note that the radii indicated as subscripts in (\ref{modspacecirc})
reflect our normalization conventions for the perturbing fields: 
Those for $J(x)$ -- dictating the radii of Dirichlet and 
Neumann circles -- are fixed by the choices in the bulk -- i.e.\ by 
$r_{\rm circ}$ --, and we have put the constant $K$ in (\ref{locmargOPE}) 
to 1 for the non-chiral deformations. 

The fibers over the bulk moduli space of orbifold models have the 
following form: 
\be
\bigl(\cM_B\bigr)_{r_{\rm orb}}\ = \ \left\{  \begin{array}{ll}
\widehat{I}_r \disju \widehat{I}_{1/2r} &\ \ r_{\rm orb} \neq \frac N{\sqrt{2}} \\[2mm] 
\hat I_r \disju \widehat{C}_{1/2r} &\ \ r_{\rm orb}= \frac N{\sqrt{2}}\,,\ 
N \geq 2  \\[2mm]
S^1_{\sqrt{2}} \disju  B_{1/\sqrt{8}} &\ \ r_{\rm orb} =  \frac 1{\sqrt{2}}  
        \end{array} \right. 
\label{modspaceorb}\ee
$\widehat{I}_r$ denotes the disjoint union of the open interval 
$\mbox{{\it \r{I}}}\,{}
=  ]0,\pi r[$ 
with four extra points for the twisted boundary states. 
The  spaces $\widehat{C}_r$ arise from the non-chiral orbifold 
deformations we did not discuss in detail above.  $\widehat{C}_r$
consists of five disjoint parts; one has the topology of an open ball 
$D^3_r \simeq \mbox{{\it \r{D}}}\,{}
^2_{\pi/\sqrt{2}} \times \mbox{{\it \r{I}}}$ 
(from the action of $\psi^a_{\rm orb}$ and $J_{\rm orb}$ 
on the untwisted Neumann boundary states), the four remaining components 
are open intervals (from  the action of $\sqrt{2}\cos(2\sqrt{2}X)$ on 
the twisted Neumann boundary states). These four intervals would form 
a single circle (and in fact do at $r_{\rm orb} = 1/\sqrt{2}$) 
were it not for the four Dirichlet-like points at which clustering 
is violated.

Some of the identifications above are as yet conjectural: Only for 
$r_{\rm circ} = \sqrt{2}$ was it possible to give precise arguments 
for the ``bagel topology'' in $\bigl(\cM_B\bigr)_m$, but it is highly 
plausible that the same picture emerges at the other exceptional radii 
$r_{\rm circ}$. The same proviso applies to the pieces  $\widehat{C}$ 
in (\ref{modspaceorb}). Also, we cannot exclude the possibility that 
there are further conformal boundary conditions at $c=1$ which 
are not continuously connected to Dirichlet or Neumann conditions 
for the current.

Except for the jumps in the fiber types occurring at multiples 
of $r=1/\sqrt{2}$, the whole space $\cM_B$ is continuous. We have indicated 
in Subsection 4.3 how to identify the fibers $S^1 \times B$ over 
$r_{\rm circ} = \sqrt{2}$ and $r_{\rm orb} = 1 / \sqrt{2}$, where 
$\cM_S^{\rm circ}$ and $\cM_S^{\rm orb}$ intersect. Over the circle 
branch, the cones describing 
Dirichlet and Neumann conditions for $r_{\rm circ} > 1 / \sqrt{2}$ 
are glued smoothly into the $S^3$ at the self-dual point. There, we 
can continuously ``change the sheet'' from Dirichlet to Neumann conditions 
for the free boson. 
\newline
This has consequences for generic radii, too. Suppose that Dirichlet 
conditions are given for a boson compactified on an arbitrary 
radius $r_{\rm circ}$. 
Combining bulk and boundary perturbations, we can continuously deform this 
situation to Neumann conditions: We first apply a marginal bulk deformation
by $J(z)\bJ(\bar z)$ until we reach the self-dual radius. There, additional 
marginal boundary fields are at our disposal to rotate the Dirichlet to Neumann 
gluing conditions on the U(1) current $J(z)$. Afterwards, $J(z)\bJ(\bar z)$ may 
shift us back to the original radius, where now Neumann conditions hold. 
The whole process never leads out the space of conformal field theories, unlike 
the D-N-transition by relevant perturbations suggested e.g.\ in \cite{FSW}. 
It shows that the dimension of a D-brane may not only change under discrete 
transformations like $T$-duality, but is not even a ``homotopy invariant'' 
for a family of boundary CFTs. 

Obviously, the moduli space of boundary conditions or 
of D-branes is much richer than that of bulk theories. In view 
of the findings of \cite{DKPS} and others that D-branes probe 
smaller distance scales in the target than strings with their 
soft scattering behaviour can do, we could say that 
``space-time'' looks richer at shorter scales. 
\newline
Let us try to explore the relation between ``space-time'' or target 
geometry and the  D-brane moduli space (\ref{modspacecirc},\ref{modspaceorb})
in more detail; after all, the study of marginal deformations should 
allow us to {\em derive} geometrical features from CFT, even when starting 
from a purely algebraic formulation of the latter. 

The $c=1$ models can be written as $\sigma$-models with $S^1$ or $S^1/\Z_2$
as the classical targets. The bulk moduli space $\cM_S$ only discloses that 
there are radii $r$ parameterizing the targets, but not their actual shape. 
It does tell us, on the other hand, that string effects induce equivalences 
between geometrically different targets: By $T$-duality, the CFT-description 
of the $\sigma$-models on $S^1_r$ and $S^1_{1/2r}$ are isomorphic, and the 
same holds for $S^1_{\sqrt{2}}$ and $S^1_{1/\sqrt{2}}/\Z_2$. 
\newline
The fibers of the brane moduli space show much more of the target 
geometry -- but still they do not simply coincide with it. Instead, 
each fiber $\bigl(\cM_B\bigr)_m$ has more connected components or even 
a higher dimension than the target corresponding to $m$. This hints at 
``non-geometric'' moduli.

Certainly, space-time supersymmetry can eliminate the corresponding   
deformations, by restricting to marginal operators which leave 
the (e.g., Dirichlet) gluing automorphism for the current intact. 
This would reinstate the standard folklore that ``the moduli space of 
BPS D0-branes is just the target of the underlying $\sigma$-model'', 
but at the cost of sweeping string-theoretic phenomena under the carpet, 
as we will see shortly. {}From a pure string world-sheet point of view, 
there is no reason anyway to discard marginal deformations that change 
$\Omega_D$. As a consequence, the very notion of world-volume dimension 
of a brane becomes ``blurred'' through (open) string effects. 

Even without invoking supersymmetry as a selection principle, 
our investigation of the $c=1$ examples suggests an interpretation 
of marginal boundary deformations that should hold in general: 
Only the operators 
present for generic values of the bulk moduli correspond to classical 
geometric moduli. At generic radius, the Dirichlet-Neumann doubling of 
the target circle or interval remains, but this is due to the discrete 
string equivalence between $T$-dual radii (note that the same 
$\cW$-automorphism governs $T$-duality and the flip of gluing conditions). 
One is inclined then to interpret any non-generic marginal perturbation 
as a signal for additional ``external'' structures like periodic tachyon 
backgrounds, which disappear as soon as an infinitesimal change in the bulk 
moduli is introduced. Sometimes, however, different such deformations 
are available, and it depends on the direction of the bulk perturbation 
which marginal boundary operators survive as ``geometric'' moduli. This 
happens at points with an ambiguous classical target interpretation of 
the bulk theory, like the meeting point of $\cM_S^{\rm circ}$ and 
$\cM_S^{\rm orb}$: While the interval swept out by the 
$\psi^2_{\rm circ}$-deformation looks non-geometric from the circle 
point of view, it is perfectly ``classical'' within the orbifold 
interpretation. 
\newline
Had we restricted ourselves to deformations which preserve the gluing 
conditions for the currents, we would have discarded $\psi^2_{\rm circ}$
from the start and would have seen no trace of the string-geometric 
identification of  $S^1_{\sqrt{2}}$ and $S^1_{1/\sqrt{2}}/\Z_2$ in 
(\ref{modspacecirc},\ref{modspaceorb}). Likewise, the ``minimal 
resolution point'' $r_{\rm circ}= 1/\sqrt{2}$ would have lost all its 
significance. If we want the D-brane moduli space to display string 
rather than classical geometry, we have to allow for seemingly 
non-geometric, gluing condition changing marginal perturbations.

There is a finer hierarchy among the ``generic'' marginal 
operators, which reflects the global symmetries of the classical 
target. Over the orbifold branch, the D-brane motion is generated by 
the ``generic'' non-chiral marginal field $J_{\rm orb}$. These deformations 
explore the underlying target even though there is no continuous
target-symmetry left after the orbifold projection from $S^1$ to 
$S^1/ \Z_2$ -- but this lack 
of symmetry becomes manifest in the partition function: The branes   
related by $J_{\rm orb}$ possess open 
string spectra which depend on the brane's distance to the 
orbifold fixed points. 
On the other hand, the periodic brane motion generated by 
$J_{\rm circ}$ (or by $J^1_{\rm orb}$ at $r_{\rm orb} = q/\sqrt{2}$) 
corresponds to a continuous target symmetry, and 
the open string spectrum is indeed invariant under the deformation. 

Note that this is just the simplest example of the (abelian or 
non-abelian) Lie group structure generally associated with chiral marginal 
perturbations. For free bosons in a torus of dimension $d\geq 1$ e.g., we 
would find subvarieties of the bulk moduli space over which the brane moduli 
space is enlarged from U(1)$^d$ to (products of) 
ADE groups of (total) rank $d$ -- see also \cite{GrGu1}. 

\smallskip

The status and the interpretation of the higher-dimensional fibers over 
exceptional points of the bulk moduli space should certainly be studied 
in more detail: 
Their topology is classical, but not group-like; they are obtained 
via an SU(2)-operation, but the matrix elements in 
(\ref{flatpertbdst},\ref{comppertbdst}) are truncated like in ``fuzzy'' 
spaces of non-commutative geometry \cite{Con} -- see also \cite{FG,FGR} 
for relations of NCG to QFT and string theory. 
\newline
Non-commutativity in brane moduli spaces was first uncovered in \cite{Wit2}; 
see also \cite{CDS,DHu}. It should be a general phenomenon occurring 
for higher central charge, connected with the interplay between marginal 
deformations and continuous parameters in the gluing conditions. 
To resolve such additional structures of the moduli space, and also 
in order to determine properties of the moduli space like 
2-body brane potentials and its metric, finer tools 
as in the exemplary treatment of \cite{Bac} will be necessary. 
In this way, it should also be possible to make contact to geometry 
and gauge theory inspired investigations of brane moduli spaces like 
e.g.\ in \cite{DGM}.  

It should not be difficult to incorporate perturbations 
by boundary condition changing operators into our analysis. The 
most prominent example where such operators occur is the condensate 
of D1-D5-strings in the D-brane derivation of the Bekenstein-Hawking 
entropy, see \cite{StVa} and also \cite{Hor,Mal1} for further details 
and references. 
\newline
Relevant boundary perturbations are important in string theory since 
they trigger the formation of D-brane bound states, see e.g.\ 
\cite{GNS,Sen1}. The CFT approach allows to study 
non-BPS bound states, too, and it was used in \cite{Sen3,Sen2,BG} 
to identify an $S$-dual pair of such states. It remains to be seen whether 
such results can be recovered  directly from relevant perturbation theory, 
by studying properties of RG-fixed points. On the other hand, very 
interesting recent work by Sen \cite{Sen4,Sen6,Sen7} shows that marginal 
boundary perturbations can often be employed as an efficient tool even 
for the study of D-brane bound states. Therefore, our general 
investigations should have applications to the $K$-theory classification 
of branes proposed in \cite{Wit4}.

As a more immediate task, the general constructions discussed in this 
paper should be applied to the 
supersymmetric case. One of the original motivations behind this work 
was to prepare the ground for a geometric interpretation of the 
Gepner model boundary states constructed in \cite{ReSc} by purely 
algebraic methods. Some promising results in this direction have been 
obtained in \cite{GuSa}, where it was also shown how the ``algebraic'' 
boundary states of \cite{ReSc} can be used to explicitly determine 
geometric quantities connected with non-perturbative D-instanton 
corrections to the moduli space geometry \cite{BBS,OOY}. We  
hope that the present methods are also 
useful in establishing further links to supersymmetric cycles 
in Calabi-Yau manifolds. 
\bigskip

{\bf Acknowledgements:} We would like to thank C.\ Bachas, I.\ Brunner,
R. Dijkgraaf, M.\ Douglas, J.\ Fr\"ohlich, K.\ Gawedzki, O.\ Grandjean,
M.\ Gutperle, N.\ Hambli, W.\ Nahm, B.\ Pioline, A.\ Sagnotti, Y.\ Satoh, 
A.\ Schwarz, A.\ Sen, Y.\ Stanev, J.-B.\ Zuber and in particular G.\ Watts 
for very useful and stimulating comments. We also benefitted from 
conversations with the participants of the DESY-workshop  on {\em Conformal 
field theory of D-branes} \footnote{for further information see 
{\tt http://www.desy.de/$\sim$jfuchs/CftD.html}} 
which provided a unique opportunity to discuss various topics 
related to this work.

\newcommand{\sbibitem}[1]{\vspace*{-1.5ex} \bibitem{#1}}

\end{document}